%% file: paper.tex
\documentclass[useAMS,usenatbib,fleqn,amssymb]{mn2e}

\usepackage{graphicx}
\usepackage{times}
\usepackage[normalem]{ulem}

\usepackage{array}
\usepackage{amssymb}
\usepackage[fleqn]{amsmath} 
\input{journal_defs}

\input{my_macros}

\def\rs{{r_{\rm s}}}
\def\rK{{r_{\rm K}}}
\def\rt{{r_{\rm t}}}
\def\rtr{{r_{\rm tr}}}
\def\rd{{r_{\rm d}}}
\def\rTF{{r_{\rm tf}}}

\title[A Tail of Two Populations]
        {A Tail of Two Populations: Chemo-dynamics of the Sagittarius Stream and Implications for Its Original Mass}
\author[Gibbons, Belokurov \& Evans]
   {S.L.J. Gibbons$^1$\thanks{E-mail:~sljg2,vasily,nwe@ast.cam.ac.uk},
    V. Belokurov$^1$ and N.W. Evans$^1$
    \medskip
  \\$^1$Institute of Astronomy, University of Cambridge, Madingley Road,
       Cambridge, CB3 0HA, UK}
\begin{document}

\date{Accepted  Received ; in original form }

\pagerange{\pageref{firstpage}--\pageref{lastpage}} \pubyear{2016}
\maketitle

\voffset-1.5cm

\label{firstpage}

\begin{abstract}
We use the SDSS/SEGUE spectroscopic sample of stars in the leading and
trailing streams of the Sagittarius (Sgr) to demonstrate the existence
of two sub-populations with distinct chemistry and kinematics. The
metallicity distribution function (MDF) of the trailing stream is
decomposed into two Gaussians describing a metal-rich sub-population
with means and dispersions $(-0.74, 0.18)$ dex and a metal-poor with
$(-1.33, 0.27)$ dex. The metal-rich sub-population has a velocity
dispersion $\sim 8$ km s$^{-1}$, whilst the metal-poor is nearly twice
as hot $\sim 13$ km s$^{-1}$. For the leading stream, the MDF is again
well-described by a superposition of two Gaussians, though somewhat
shifted as compared to the trailing stream. The metal-rich has mean
and dispersion $(-1.00, 0.34)$ dex, the metal-poor $(-1.39, 0.22)$
dex. The velocity dispersions are inflated by projection effects to
give $15$ - $30$ kms$^{-1}$ for the metal-poor, and $6$ - $20 $
kms$^{-1}$ for the metal-rich, depending on longitude. We infer that,
like many dwarf spheroidals, the Sgr progenitor possessed a more
extended, metal-poor stellar component and less extended, metal-rich
one. We study the implications of this result for the progenitor mass
by simulating the disruption of the Sgr, represented as King light
profiles in dark halos of masses between $10^{10}$ and $10^{11}
M_\odot$, in a three-component Milky Way whose halo is a live
Truncated Flat potential in the first phase of accretion and a
triaxial Law \& Majewski model in the second phase.  We show that that
the dark halo of the Sgr must have been $\gtrsim 6 \times 10^{10}
M_\odot$ to reproduce the run of velocity dispersion with longitude
for the metal-rich and metal-poor sub-populations in the tails.
\end{abstract}


\section{Introduction}\label{sec:introduction}

The disruption of the third most massive satellite of the Galaxy, the
Sagittarius dwarf (Sgr) is currently in full swing. Its tidal debris
has been sprayed all over the sky in a theatrical demonstration of the
messy eating habits of the more massive Milky Way (MW)
Galaxy. Exemplifying the hierarchical nature of structure assembly on
small scales, the Sgr accretion event has been used to champion the
$\Lambda$CDM theory. Over the last decade, its tidal tails have been
mapped across a large Galactic volume and have been utilized to study
the matter distribution in the Milky Way. However, there remain
embarrassing lacunae that threaten to mar the stream based inference
procured so far: we know neither the Sgr's original mass nor where the
dwarf came from.

In fact, as shown by \citet[][]{jiang2000}, the Sgr mass and the
initial apocentric distance of its orbit are tightly linked. The
current position of the remnant can be reproduced by a large range of
initial conditions from low apocentres at around 60 kpc to those
beyond the virial radius of the Milky Way, i.e., $>200$ kpc. On one end
of the spectrum, low size orbits are allowed to exist coupled with a
light Sgr dwarf with $M\sim10^9M_{\odot}$; on the other hand, a
substantially heavier Sgr with $M\sim10^{11}M_{\odot}$ could plausibly
start much further away, but then sink quickly into the central Galaxy
thanks to dynamical friction with the MW's dark halo. Thus, the
dynamical modeling permits an uncertainty of two orders of magnitude
in mass. Unfortunately, this cannot be improved through direct
observation: the original properties of the dwarf and its orbit are
simply not available today. There are however indirect indications that
can be invoked to help deduce what the progenitor's mass might have
been.

\begin{figure*}
\centering
\includegraphics[width=0.33\textwidth]{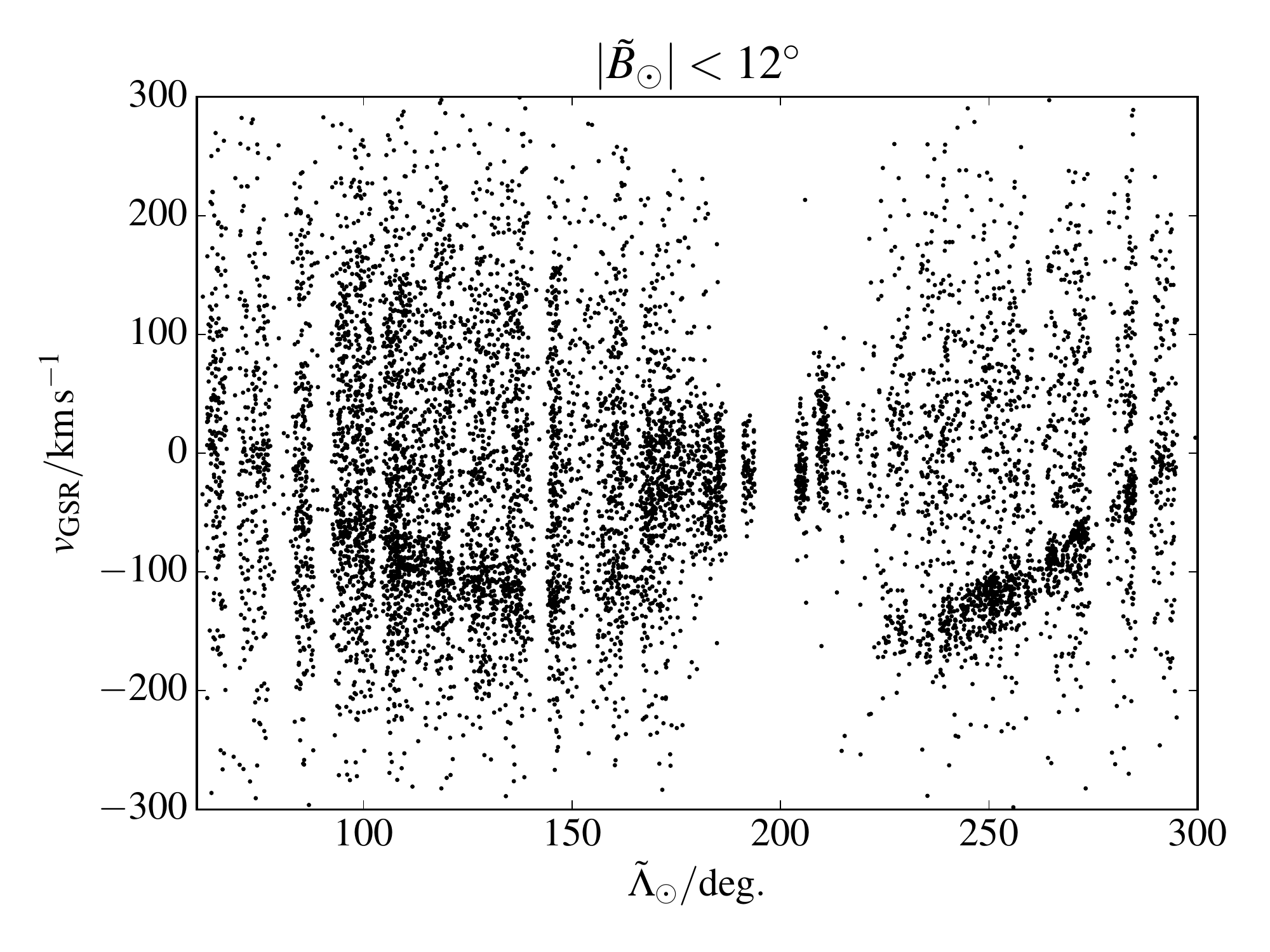}
\includegraphics[width=0.33\textwidth]{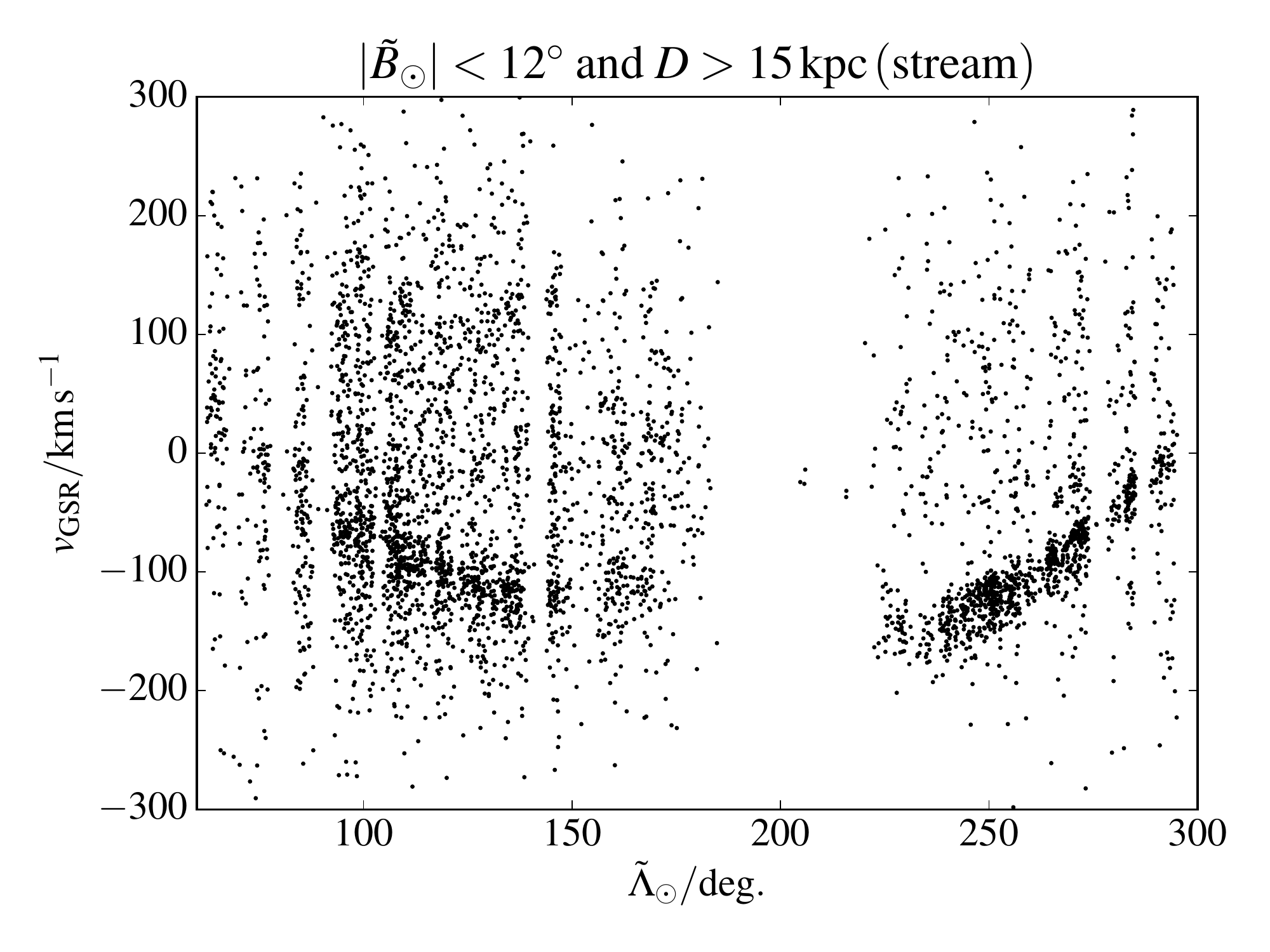}
\includegraphics[width=0.33\textwidth]{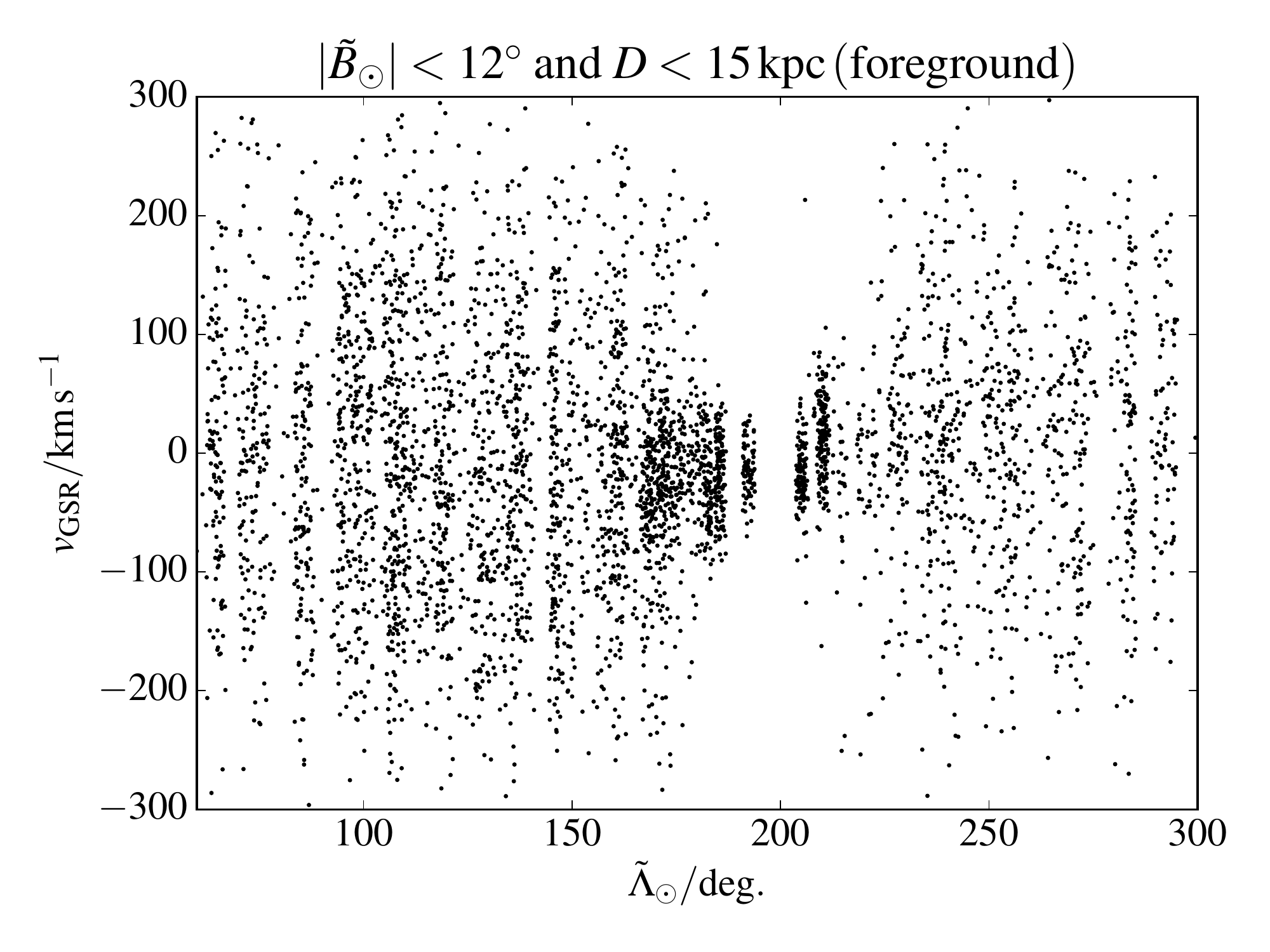}
\caption{Our sample of SDSS/SEGUE stars focused on the Sagittarius
  Stream in the $v_{\rm GSR}$ versus $\tilde{\Lambda}_\odot$ plane. We
  show the selection before (left panel) and after (middle panel) the
  application of the distance cut, as well as the stars removed by the
  distance cut (right panel).  The leading stream is visible as the
  overdensity with $\tilde{\Lambda}_\odot < 150^\circ$. The trailing
  stream is the overdensity with $\tilde{\Lambda}_\odot >
  220^\circ$. The distance cut significantly reduces Milky Way
  contamination, whilst leaving the stream signal intact.  }
\label{fig:vgsr_selection}
\end{figure*}

\citet{LM2010} attempt to address the Sgr mass conundrum by taking
advantage of the measurement of the velocity dispersion of the
trailing tail debris. They conjecture that a more massive progenitor
ought to host a stellar population with a higher intrinsic velocity
dispersion that would produce a hotter tidal stream on disruption -- a
hypothesis they convincingly prove to be correct with a suite of
numerical simulations of the Sgr disruption. Note that only a simple
one-component (representing both DM and stars) model for the
progenitor was used. Nonetheless, encouraged by the clear correlation
between the stream's dispersion and the progenitor's mass, they relied
on the most precise measurement of the Sgr debris dispersion available
at the time. This was established by \citet[][]{Monaco2007}, who used
high-resolution spectroscopy of the Sgr trailing tail giant stars to
infer the dispersion of $8.3 \pm 0.9$ km s$^{-1}$. Given the ensemble
of models produced by \citet[see e.g.][]{LM2010}, the corresponding
Sgr progenitor's mass is $6.4 \times 10^8 M_{\odot}$, assuming the
disruption duration of 8 Gyr.

There are (at least) three serious problems with such a low estimate
of the original mass of the dwarf. First, as implied by the analysis
of \citet[][]{jiang2000}, such a light Sgr does not have enough mass to
experience any significant dynamical friction during its evolution in
the MW potential. Therefore, it must have started its orbit at around
60 kpc from the Galactic centre, i.e. around its current
apocentre. Note that the present day apocentre is fixed by the
precise knowledge of the apocentres of the leading and trailing tails
\citep[see e.g.][]{Belokurov2014}. Cosmologically speaking, this makes
little sense. The accretion of Sgr is a relatively recent event,
dating back only to approximately $z\sim0.6$. The most compelling
evidence in this regard comes from the studies of the dwarf's stellar
populations. An approximate scale for the time elapsed since the
beginning of the disruption is set by the M giants found in the Sgr
trailing tail: their age is bracketed to lie between 4 and 9 Gyr
\citep[see e.g.][]{Bellazini2006}. This is in agreement with
\citet{deBoer2015} who demonstrate that the star formation activity in
the stream ceased abruptly between 5 and 7 Gyr ago. Since $z\sim0.6$,
the Milky Way's DM mass within the virial radius has experienced only
a minuscule increase \citep[see e.g.][]{Diemer2013, Wetzel2015}, thus
rendering the turn-around radius of 60 kpc implausible.

\begin{figure}
\centering
\includegraphics[width=0.4\textwidth]{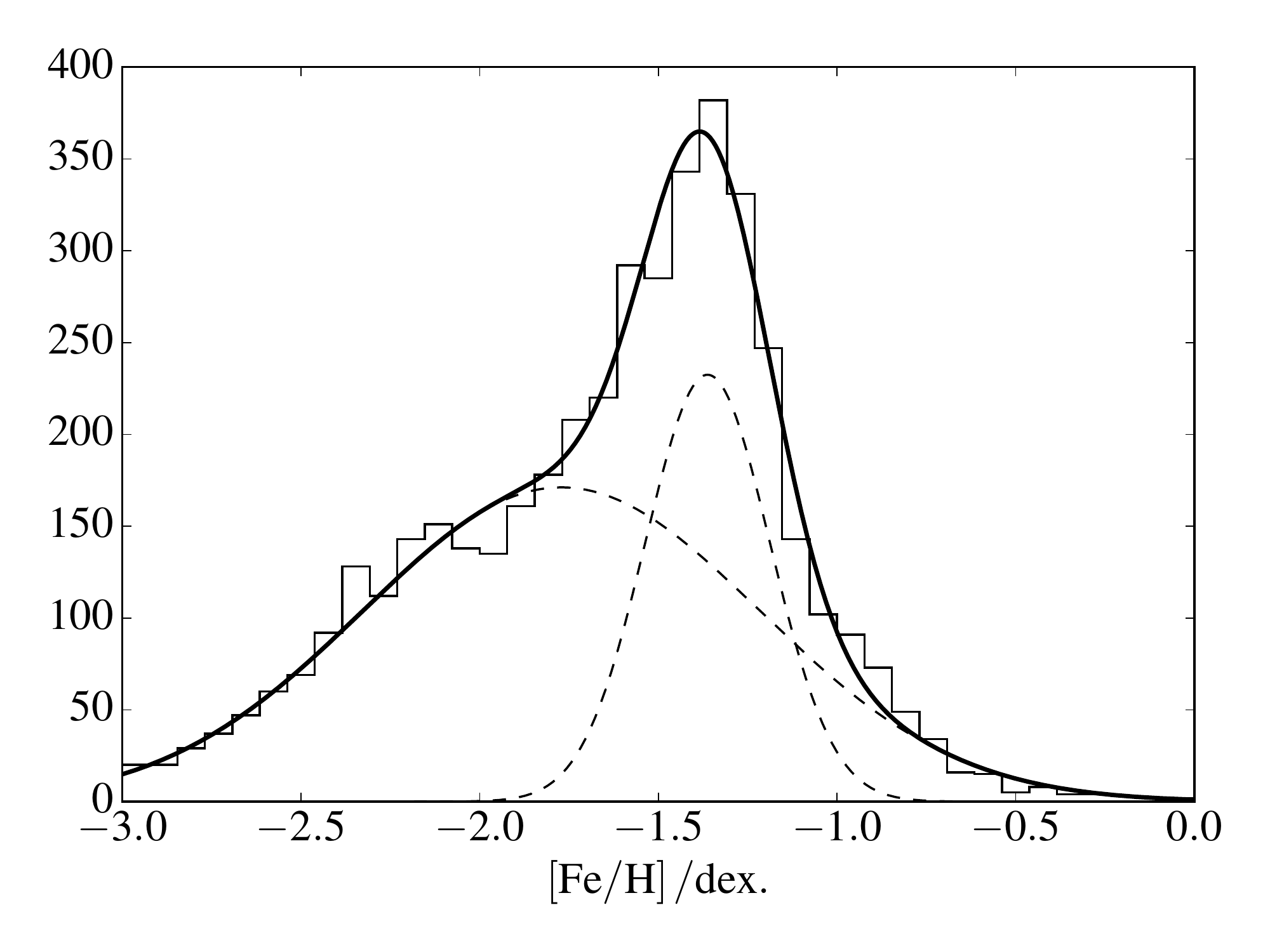}
\caption{The MDF of all SDSS/SEGUE stars satisfying our selection that
  are located off the Sagittarius Stream ($|\tilde{B}|_\odot \geq
  15^\circ$). It is well fit by two Gaussians (solid black curve)
  with components indicated by the dashed black lines.}
\label{fig:halo_mdf}
\end{figure}

Furthermore, the progenitor's mass of $\sim6\times10^8 M_{\odot}$
appears inconsistent with the most recent estimate of the remnant's
current mass. Through modeling of a large sample of high-resolution
NIR spectra as part of the APOGEE survey, \citet[][]{Majewski2013}
have gauged the remnant's mass to be in the range $5-7 \times 10^8
M_{\odot}$. In other words, after 6-8 Gyr of continuous tidal
stripping, the remnant's mass today is claimed to be of order of the
original progenitor's mass! Finally, the dwarf's mass before
disruption can be guessed if its original luminosity was
known. \citet[][]{Martin2010} provide exactly such an estimate. By
carefully unpicking the Sgr tidal debris from the foreground stellar
populations across the entire sky, they calculate the lower limit to
the total stellar mass in the Sgr progenitor. According to
\citet[][]{Martin2010} and considering the analysis presented in
\citet[][]{Martin2012}, Sgr at infall must have contained $\sim1.4
\times 10^8 M_{\odot}$ in stars alone. Guided by the abundance
matching relations \citep[see e.g.][]{Conroy2009,Behroozi2010}, the
corresponding halo mass should be of order of $10^{11} M_{\odot}$.

Taking into account the arguments above, it is more likely that the Sgr
dwarf's original mass was much larger, i.e. closer to $10^{11}
M_{\odot}$. Such a heavy Sgr would of course wreak havoc in the Milky
Way's disk. As pointed out by \citet[][]{jiang2000}, the Galaxy's HI
layer should show clear signs of interaction with the most massive of
their versions of the dwarf. Naturally, the disk's stellar component
is likely to be disturbed just as well. Observationally, there are,
actually, signs of such disturbances in both the neutral hydrogen
layer -- as represented by the spiral arms and the warp -- and the
stellar disk, as evidenced by the discovery of the wave-like
oscillations in the Sloan Digital Sky Survey (SDSS) star counts
\citep[][]{Widrow2012,Yanny2013}. The link between the Sgr accretion
and sub-structures in the disk has been recently firmed up through
numerical simulations. For example, \citet[][]{Purcell2011} establish
connection between the Sgr infall and the Galactic spiral structure,
while \citet{Gomez2016} identify the Sgr dwarf as their likely
candidate to kick out large numbers of disk stars to form Monoceros
ring-like structures.

A Sgr progenitor significantly more massive than that proposed by
\citet[][]{LM2010} would superficially be in tension with the low
velocity dispersion measured by \citet[][]{Monaco2007} in the trailing
tail. A fresh look at the kinematics of the trailing debris is
presented in \citet[][]{Koposov2013}, who measure $\sim14 \pm 1$ km
s$^{-1}$ only a few degrees away from the location where
\citet[][]{Monaco2007} obtain $8.3 \pm 0.9$ km s$^{-1}$. These two
measurements are not consistent with each other at $>3 \sigma$
level. A possible resolution of this apparent inconsistency can be
found in the analysis of \citet[][]{Majewski2013}, who discover two
distinct stellar populations in the Sgr remnant. According to their
Figure~2, there exists a metal-poor sub-population with typical
velocity dispersion of $\sim14$ km s$^{-1}$ and a metal-rich
sub-population with a dispersion of $\sim 8$ km s$^{-1}$. Therefore,
we hypothesize that a similar stellar population dichotomy might exist
in the stream itself. Then, the difference between the dispersion
measurements of \citet[][]{Monaco2007} and \citet[][]{Koposov2013} can
be explained away if at least one of the spectroscopic samples
displayed a strong metallicity bias. A quick look at the Table 5 and
the lower panel of Figure 9 in \citet[][]{Monaco2007} reveals that,
indeed, their measurements are biased towards the highest metallicity
members (M-giant stars) of the trailing tail. This is contrast to the
SDSS spectroscopic sample used by \citet[][]{Koposov2013}, where the
metallicity distribution is substantially broader, encompassing the
range between [Fe/H]=-3 and [Fe/H]=-0.5.

\begin{figure}
\centering
\includegraphics[width=0.33\textwidth]{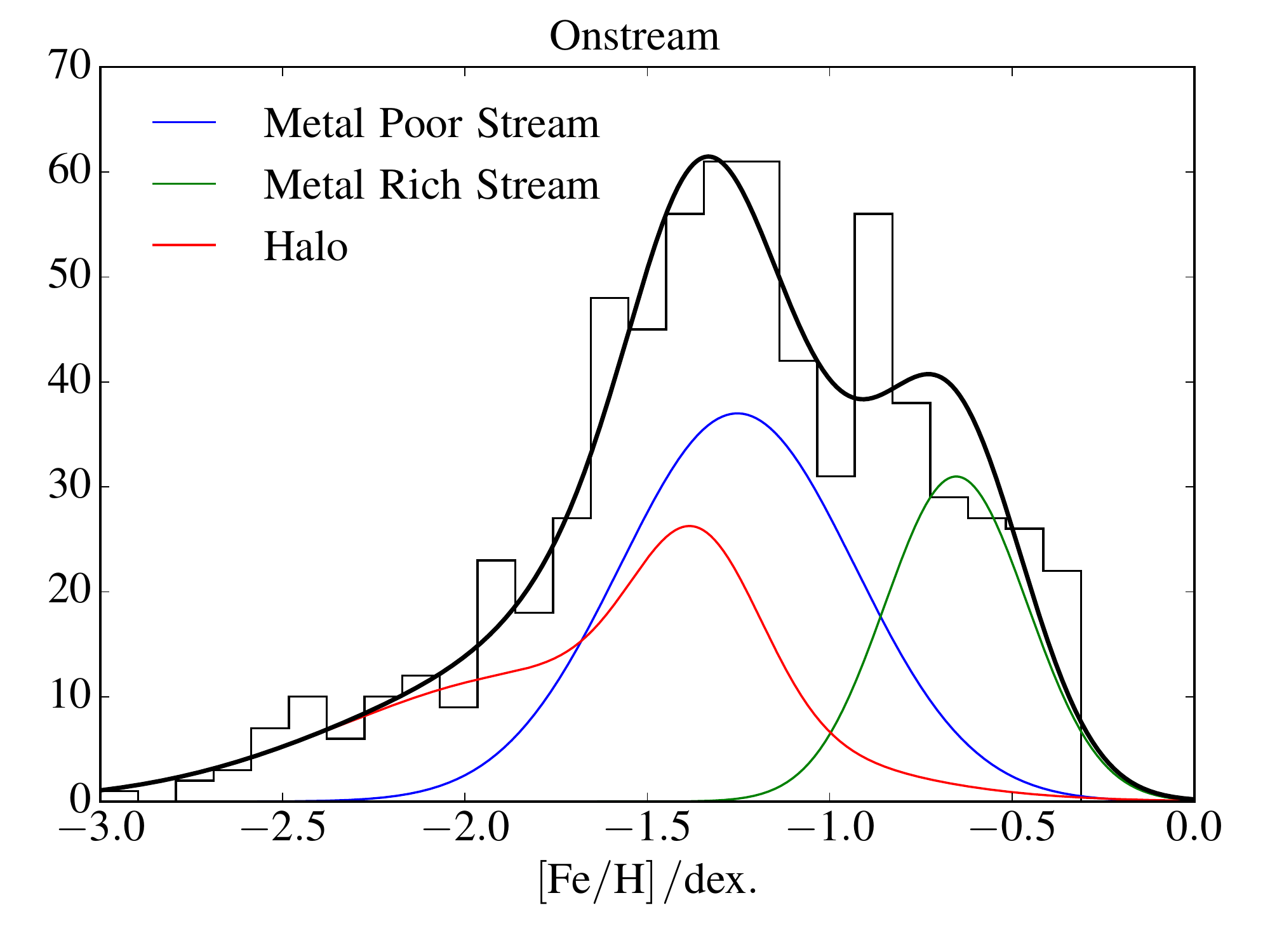}\\
\includegraphics[width=0.33\textwidth]{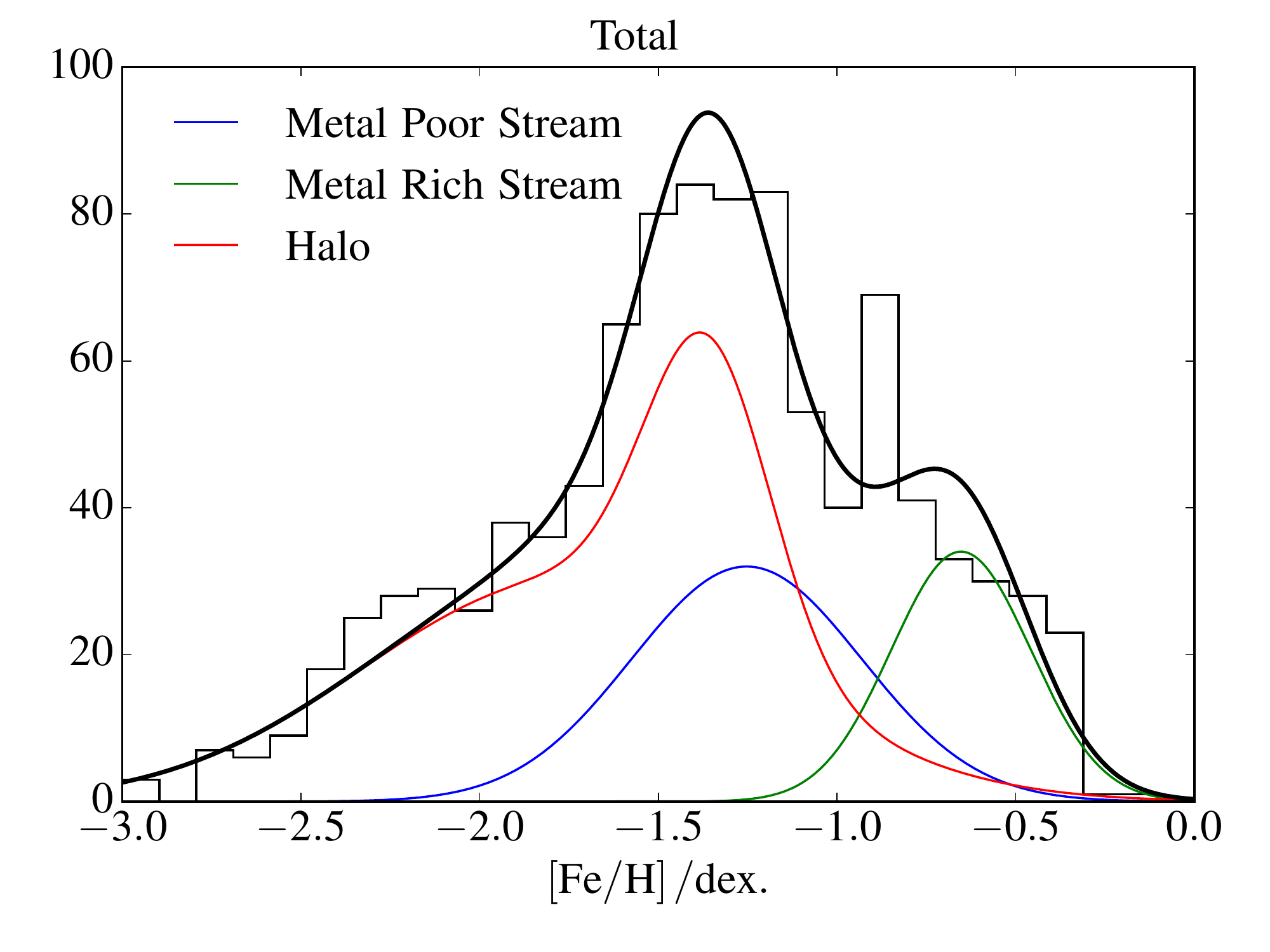}
\caption{The upper panel shows the MDF of the stars in the trailing
  stream, with $235^\circ \leq \tilde{\Lambda}_\odot \leq 274^\circ$,
  and velocities within $\pm45~\kms$ of the mean stream velocity. The
  green and blue lines show the metal-rich and metal-poor Sgr
  components, while the red line shows the contribution from the Milky
  Way component. The lower panel shows all the stars within the
  aforementioned range of $\tilde{\Lambda}_\odot$, with the components
  rescaled.}
\label{fig:mdf_trailing}
\end{figure}

Encouraged by the detection of multiple stellar populations in the Sgr
remnant, we present here an in-depth study of the chemo-dynamical
properties of the Sgr tidal tails. More precisely, we model the
spectroscopy of the likely Sgr stream members available as part of
SDSS DR9 (Section~\ref{sec:spec}). According to our analysis, across
the entire SDSS footprint, stars in both the leading and the trailing
tail exhibit a clear dichotomy, in the sense that the more metal-rich
population possesses a lower line-of-sight velocity dispersion in
comparison to the more metal-poor one. Our findings immediately remove
any seeming tension between the results of \citet[][]{Monaco2007} and
\citet[][]{Koposov2013}. Naturally, the presence of a substantially
hotter metal-poor population in the stream demands a more massive
progenitor than that inferred by \citet[][]{LM2010}. The link between
the stream dispersion and the progenitor mass can be established
through an analysis of an extensive suite of numerical simulations,
such that the effects of the dynamical friction are taken into
account. Such comprehensive analysis is beyond the scope of this
paper. However, to gauge the range of the Sgr progenitor masses
allowed by our dispersion measurements, we carry out a series of
N-body experiments where two-component Sgr dwarf prototypes are
accreted by realistic MW hosts in the presence of dynamical friction
(Section~\ref{sec:sims}). The implications of our investigation are
reported in Section~\ref{sec:results}.

\begin{figure}
\centering
\includegraphics[width=0.33\textwidth]{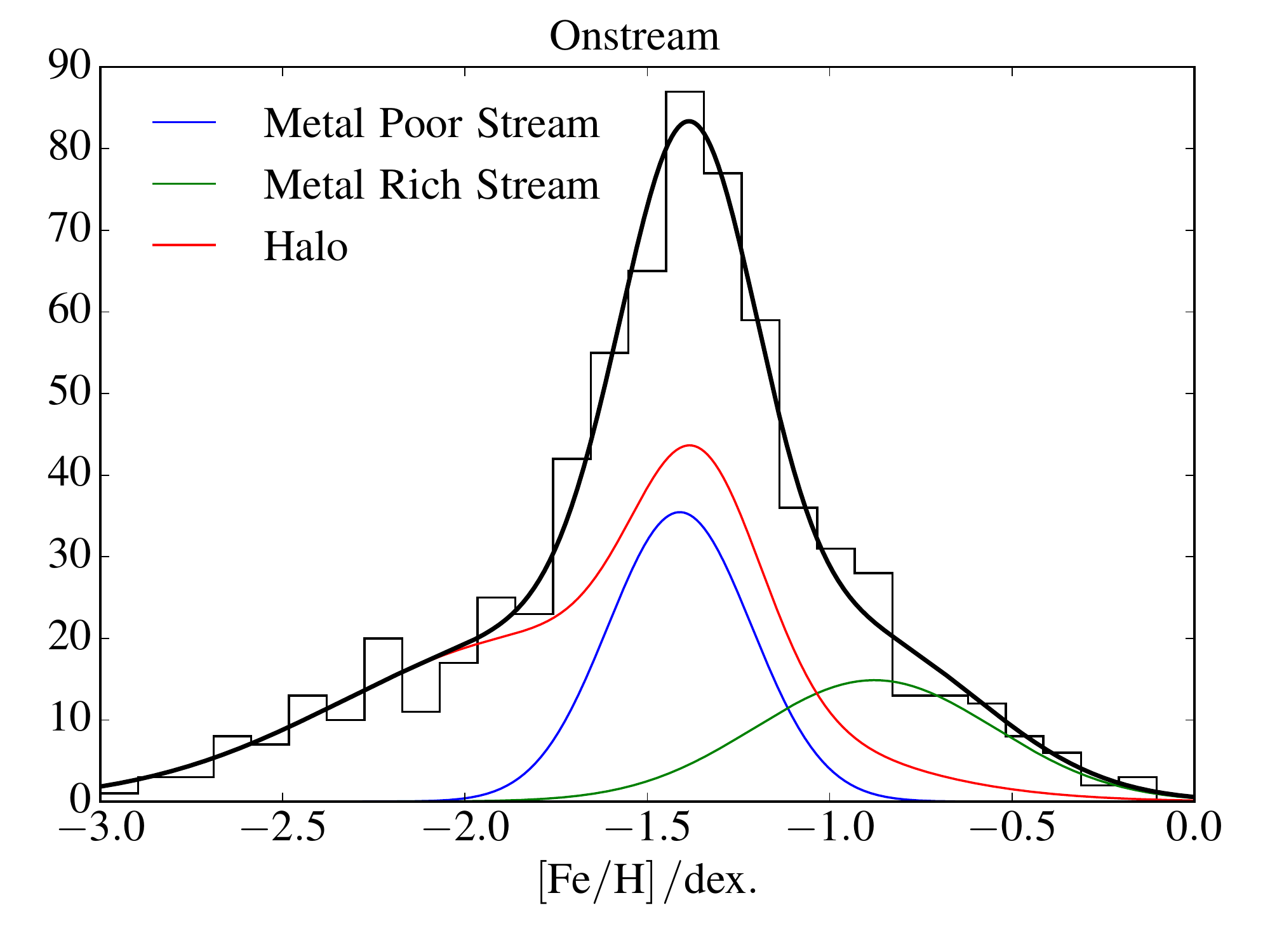}\\
\includegraphics[width=0.33\textwidth]{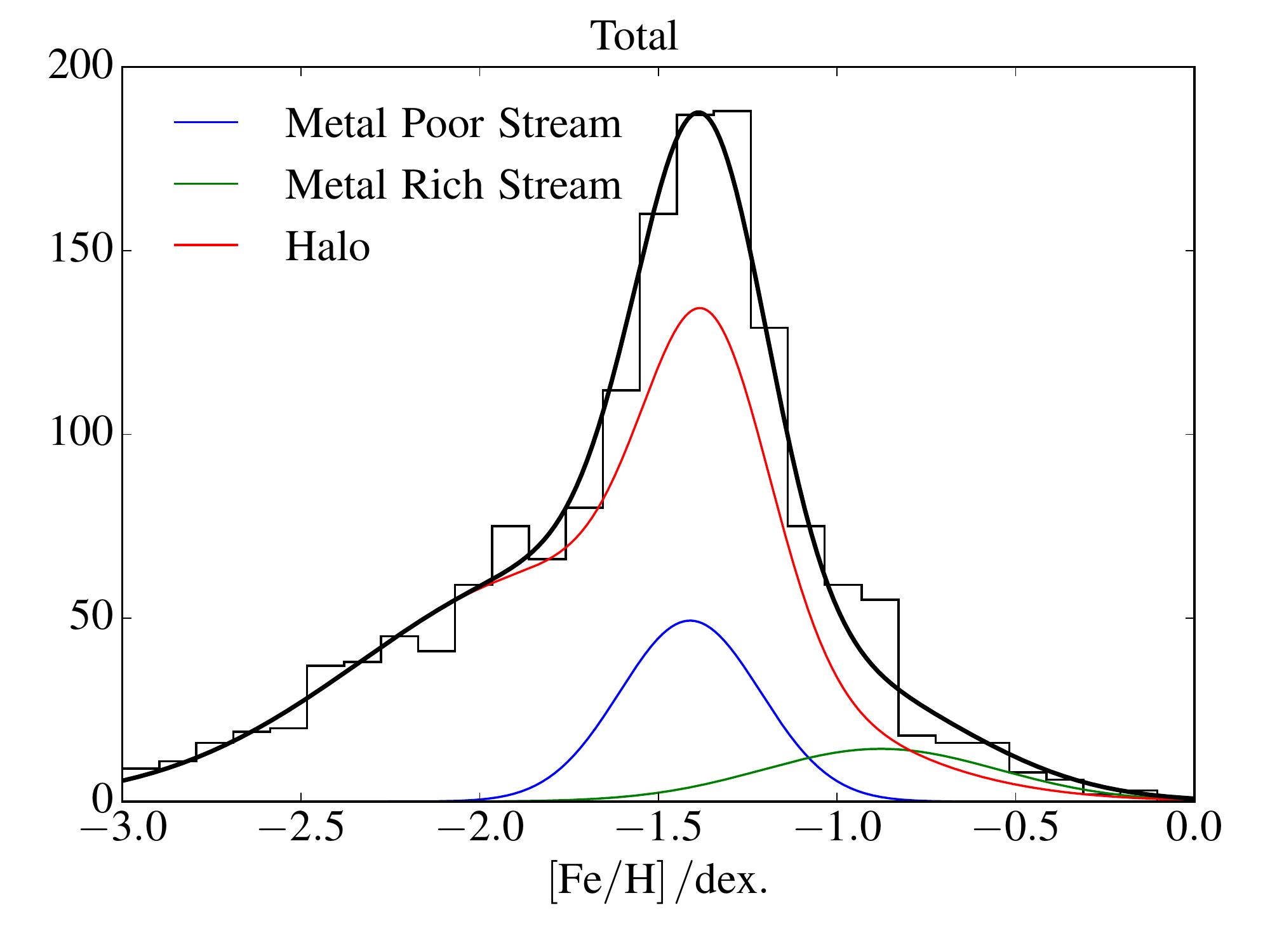}
\caption{As Fig.~\ref{fig:mdf_trailing}, but for the leading stream
  stars with $\tilde{\Lambda}_\odot < 150^\circ$.}
\label{fig:mdf_leading}
\end{figure}

\section{Chemistry and Kinematics of the Sgr Stream in the SDSS DR9}
\label{sec:spec}

\subsection{Spectroscopic Selection} \label{sec:spec_sel}

\begin{table*}
\centering
\begin{tabular}{l | c c c c}
&$234^\circ < \tilde{\Lambda}_\odot <248^\circ$&$248^\circ < \tilde{\Lambda}_\odot <255^\circ$&$255^\circ < \tilde{\Lambda}_\odot <266^\circ$&$266^\circ < \tilde{\Lambda}_\odot <274^\circ$\\
\hline
$N_{\rm stars}$&257&242&228&254\\
\hline
$w_1$&$0.473\pm^{0.040}_{0.038}$&$0.478\pm^{0.039}_{0.040}$&$0.429\pm^{0.040}_{0.040}$&$0.390\pm^{0.040}_{0.041}$\\
$w_2$&$0.141\pm^{0.029}_{0.027}$&$0.156\pm^{0.030}_{0.028}$&$0.154\pm^{0.030}_{0.028}$&$0.089\pm^{0.026}_{0.022}$\\
\hline
$\sigma_1 \, / \kms$&$15.7\pm^{1.5}_{1.3}$&$13.5\pm^{1.2}_{1.0}$&$12.5\pm^{1.2}_{1.0}$&$12.6\pm^{1.5}_{1.5}$\\
$\sigma_2 / \kms$&$14.0\pm^{2.7}_{2.6}$&$8.5\pm^{1.5}_{1.2}$&$7.1\pm^{1.7}_{1.3}$&$6.4\pm^{3.8}_{2.6}$\\
$\sigma_{MW} / \kms$&$113.6\pm^{8.9}_{7.9}$&$114.7\pm^{9.3}_{8.6}$&$106.8\pm^{8.1}_{7.0}$&$105.3\pm^{6.8}_{6.2}$\\
\hline
$\bar{v_1} \, / \kms$&$-142.5\pm^{1.9}_{1.9}$&$-124.2\pm^{1.2}_{1.2}$&$-107.4\pm^{1.3}_{1.3}$&$-79.0\pm^{1.5}_{1.5}$\\
$\dd \bar{v_1} / \dd \Lambda  / \, \kms \, {\rm deg.}^{-1}$&$2.24\pm^{0.36}_{0.37}$&$0.75\pm^{0.47}_{0.49}$&$2.59\pm^{0.29}_{0.29}$&$3.53\pm^{0.54}_{0.53}$\\
$\bar{v_2} \, / \kms$&$-143.5\pm^{3.8}_{4.0}$&$-114.7\pm^{1.6}_{1.7}$&$-98.6\pm^{1.5}_{1.5}$&$-75.7\pm^{1.6}_{2.2}$\\
$\dd \bar{v_2} / \dd \Lambda /  \, \kms \, {\rm deg.}^{-1}$&$3.97\pm^{0.71}_{0.69}$&$0.30\pm^{0.63}_{0.64}$&$2.74\pm^{0.34}_{0.35}$&$1.97\pm^{0.49}_{0.59}$\\
\hline
\end{tabular}
\caption{Fitted parameters for the trailing stream, including the
  number of stars in each bin $N$, the weights of the Gaussians $w_i$,
  and the velocity dispersions $\sigma_i$ around the mean velocity
  tracks for the metal-rich and metal-poor populations. }
\label{tab:model_trailing}
\end{table*}
\begin{table*}
\centering
\begin{tabular}{l | c c c c}
&$100^\circ < \tilde{\Lambda}_\odot <110^\circ$&$110^\circ < \tilde{\Lambda}_\odot <120^\circ$&$120^\circ < \tilde{\Lambda}_\odot <130^\circ$&$130^\circ < \tilde{\Lambda}_\odot <140^\circ$\\
\hline
$N_{\rm stars}$&347&328&420&455\\
\hline
$w_1$&$0.181\pm^{0.032}_{0.031}$&$0.143\pm^{0.030}_{0.028}$&$0.152\pm^{0.030}_{0.029}$&$0.160\pm^{0.033}_{0.029}$\\
$w_2$&$0.088\pm^{0.023}_{0.020}$&$0.135\pm^{0.024}_{0.022}$&$0.106\pm^{0.025}_{0.022}$&$0.121\pm^{0.027}_{0.024}$\\
\hline
$\sigma_1 \, / \kms$&$31.4\pm^{7.7}_{5.9}$&$21.6\pm^{4.5}_{3.5}$&$15.0\pm^{4.6}_{3.2}$&$16.9\pm^{3.5}_{3.0}$\\
$\sigma_2 / \kms$&$19.7\pm^{6.5}_{4.3}$&$12.0\pm^{1.9}_{1.5}$&$6.0\pm^{1.9}_{1.4}$&$10.5\pm^{2.0}_{1.7}$\\
$\sigma_{MW} / \kms$&$110.3\pm^{4.7}_{4.4}$&$115.7\pm^{4.9}_{4.5}$&$115.8\pm^{5.6}_{5.1}$&$127.9\pm^{6.0}_{5.6}$\\
\hline
$\bar{v_1} \, / \kms$&$-87.6\pm^{6.6}_{7.3}$&$-107.2\pm^{3.8}_{3.9}$&$-110.7\pm^{3.0}_{3.5}$&$-123.8\pm^{4.0}_{3.8}$\\
$\dd \bar{v_1} / \dd \Lambda  / \, \kms \, {\rm deg.}^{-1}$&$-2.3\pm^{1.6}_{1.6}$&$-1.59\pm^{0.91}_{0.93}$&$0.89\pm^{0.80}_{0.84}$&$-3.5\pm^{1.3}_{1.2}$\\
$\bar{v_2} \, / \kms$&$-77.5\pm^{6.2}_{5.9}$&$-90.2\pm^{1.9}_{2.0}$&$-107.1\pm^{1.3}_{1.3}$&$-116.7\pm^{2.5}_{2.6}$\\
$\dd \bar{v_2} / \dd \Lambda /  \, \kms \, {\rm deg.}^{-1}$&$-1.9\pm^{1.3}_{1.4}$&$-0.65\pm^{0.50}_{0.54}$&$-2.72\pm^{0.50}_{0.47}$&$0.79\pm^{0.70}_{0.74}$\\
\hline
\end{tabular}
\caption{As Table~\ref{tab:model_trailing}, but for the leading
  stream.}
\label{tab:model_leading}
\end{table*}

To study the chemistry and the kinematics of the Sagittarius stream we
take advantage of the largest spectroscopic dataset to date - the
SDSS. More precisely, we use the combination of the SDSS and SEGUE
spectroscopic surveys, available in SDSS DR9 \citep{SDSSDR9,
  Yanny2009}. This consists of medium-resolution spectroscopy of stars
covering a large portion of the sky, with many fields overlapping with the
leading and trailing arms of the Sagittarius stream. The stellar
parameters and metallicities of individual stars were derived through
the detailed fitting of synthetic spectra using the SEGUE Stellar
Parameter Pipeline \citep[see e.g.][]{Allende2008, Lee2008,
  Lee2011, Smolinski2011}.

As a first step, we identify the likely stream member stars by
transforming the equatorial Right Ascension and Declination
coordinates to the heliocentric stream-aligned positions
$(\tilde{\Lambda}_\odot, \tilde{B}_\odot)$, as defined in
\citet{Majewski2003}.  Note that following the convention of
\citet{Belokurov2014}, $\tilde{\Lambda}_\odot$ increases in the
direction of the Sgr progenitor's motion and $\tilde{B}_\odot$ points
towards the Galactic North Pole. The stream member stars are selected
as all spectroscopic targets with $|\tilde{B}_\odot| \leq 12^\circ$.
To ensure that all targets have robust stellar parameters, a minimum
signal-to-noise ratio of 25 is enforced. Finally, following
\citet{deBoer2015}, we further refine the selection by choosing the
spectroscopically confirmed giants. These are selected as stars with
$\log g \leq 3.5$ and $4300 \leq T_{\rm eff} \leq 6000 \K$. The
combination of the selection criteria described above yields
$\sim$9,000 stars in total.

Fig.~\ref{fig:vgsr_selection} shows the distribution of the selected
stars in the plane of the radial velocity (corrected for the Solar
reflex motion) $v_{\rm GSR}$ and the Sgr stream longitude
$\tilde{\Lambda}_\odot$ (left panel). Clearly, even with the above
cuts in place, the effect of the Galactic contamination is still
appreciable, especially for the leading stream in the North. The
foreground populations come from both the nearby halo, as evidenced by
the large velocity dispersion across the whole range of
$\tilde{\Lambda}_\odot$, as well as the thick disk as revealed by a
portion of stars following a narrow velocity track at low Galactic
latitudes (e.g. $150^{\circ} < \tilde{\Lambda}_\odot <
230^{\circ}$). To reduce the foreground presence, we utilize the fact
that the stream in these parts of the sky is at a much greater
distance, i.e. $(>15\,\kpc)$ than the bulk of the Milky Way disk and
halo stars \citep{Belokurov2014}. Therefore, we proceed as follows. We
derive distances for the stars in our sample by placing them on the
PARSEC isochrones \citep{Bressan2012}. The isochrone's metallicity and
surface gravity are chosen according to the star's metallicity [Fe/H]
and surface gravity $\log g$ derived from spectroscopy, while the age
is kept constant at $8\times10^9$ yr. With these assumptions, the
stellar absolute magnitudes are determined in the $g$, $r$, $i$ and
$z$ bands. The median of the four absolute magnitude estimates is
taken as our measured distance. Accordingly, all stars with a derived
distance in excess of $15\kpc$ are removed from the sample. This
approach is similar in spirit to the de-contamination procedure
employed by \citet{deBoer2015}, though differing in detail.

The final (cleaned) selection of $\sim$4,300 stars is shown in the
middle panel of Fig.~\ref{fig:vgsr_selection}. For comparison, the
stars rejected by the distance cut are displayed in the right panel of
the Figure.  Reassuringly, it can be seen that the method
substantially reduces the contamination from the Milky Way disk+halo
whilst leaving the stream signal intact. In fact, it appears that the
majority of the remaining MW population is that of the halo, as no
clear velocity gradient is visible - other then that of the Sgr stream
itself - in the middle panel of the Figure. Note that, even after the
substantial cleaning of the sample, the leading tail, i.e. stars with
$\tilde{\Lambda}_\odot < 200^{\circ}$ appears to be more affected by
the Galactic contamination as compared to the trailing debris,
i.e. stars with $\tilde{\Lambda}_\odot > 200^{\circ}$.

\subsection{Metallicity Distribution Function of the Galactic halo and the Sgr Stream}

Before the behaviour of the stream can be studied in the space of
metallicity and velocity, a model for the distribution of the Milky
Way stars must be established. To this end, we empirically determine
the metallicity distribution function (MDF) of the Milky Way
contaminant population by selecting all stars satisfying the criteria
described in Section~\ref{sec:spec_sel}, but with $|\tilde{B}_\odot|
\geq 15^\circ$ to exclude the majority of stream members. The MDF
returned by this selection is displayed as a histogram in
Fig.~\ref{fig:halo_mdf}. We find that this metallicity distribution
can be represented adequately with a mixture of two Gaussians
(individual components shown as dashed lines, and their sum as a solid
line). Of course, this distribution is obviously not the \emph{true}
underlying MDF of the Milky Way halo -- but is, instead, the
distribution modulated by the SEGUE selection function. The two
Gaussians in our model (determined through maximum likelihood analysis
of the SDSS spectra) have the following centers and widths:
$(\mu^{[Fe/H]}_{MW1}, \sigma^{[Fe/H]}_{MW1})=(-1.38, 0.15)$ dex and
$(\mu^{[Fe/H]}_{MW2}, \sigma^{[Fe/H]}_{MW2})=(-1.81,0.56)$ dex. Even
while affected by the selection bias, this blend of a narrow
metal-rich ingredient and a broad metal-poor is not inconsistent with
the chemical properties of the Galactic halo in the literature
\citep[see e.g.][]{Helmi2008, Beers2012,deepsloanspec}.

To measure the MDF of the stream we exploit the fact that it displays
a clear trend in velocity $v_{\rm GSR}$ as a function of
$\tilde{\Lambda}_\odot$ longitude. To boost the stream signal and to
minimize the Milky Way contamination, we select stars with velocities
within $\pm 45 \kms$ from the mean stream track in $v_{\rm GSR}$
space, as determined in \citet{Belokurov2014}.  The half-width of the
selection region in the phase space corresponds to approximately
$\pm3\sigma$ of the velocity dispersion found in
\citet{Koposov2013}. Fig.~\ref{fig:mdf_trailing} shows the MDF for the
trailing tail, represented by the Sgr candidate stars with $234^\circ
< \tilde{\Lambda}_\odot < 270^\circ$. The top panel displays only
stars within the velocity range of the mean track of the stream, while
the bottom panel includes all stars in the this range of
$\tilde{\Lambda}_\odot$. As expected, the only difference between the
two panels is the relative scaling of the stream and the halo.

The metallicity distribution displayed in
Figure~\ref{fig:mdf_trailing} is markedly bi-modal, thus indicating
two distinct populations of stars in the stream. Motivated by this
discovery, we model the distribution with a mixture of the contaminant
MW MDF and two additional Sgr trailing (ST) Gaussians with
$(\mu^{[Fe/H]}_{\rm ST1},\sigma^{[Fe/H]}_{\rm ST1})=(-1.33, 0.27)$ and
$(\mu^{[Fe/H]}_{\rm ST2},\sigma^{[Fe/H]}_{\rm ST2})=(-0.74, 0.18)$
dex. Compared to the halo overall, the stream lacks a substantial
metal-poor component. Instead, there is a conspicuous metal-rich
population, in agreement with a multitude of previous studies of both
the remnant and the stream \citep[see e.g.][]{Sbordone2007, YannySgr,
  Chou2010, Koposov2013, deBoer2014, deBoer2015}

Similarly, Fig.~\ref{fig:mdf_leading} presents the MDF of the leading
tail, i.e. the portion of the stream with $\tilde{\Lambda}_\odot <
150^\circ$.  Here, the metal-rich sub-population is less obvious
compared to the trailing stream. Through a maximum-likelihood analysis,
we find that a combination of these two Gaussians describe the Sgr
leading (SL) MDF well: $(\mu^{[Fe/H]}_{\rm SL1},\sigma^{[Fe/H]}_{\rm
  SL1})=(-1.39, 0.22)$ and $(\mu^{[Fe/H]}_{\rm
  SL2},\sigma^{[Fe/H]}_{\rm SL2})=(-1.00, 0.34)$. In the leading tail
MDF, the centres of the metallicity distributions of the two
sub-populations are closer to each other, with a more significant
overlap as compared to those in the trailing tail. Also, note that the
metal-rich component of the MW foreground model is coincident with the
metal-poor sub-population of the Sgr leading tail.

These differences and similarities between the MDFs of the two tails
and the MW halo reflect various astrophysical and observational
effects at play. To begin with, there exists a known metallicity
gradient along the Sgr stream, and the regions probed by our analysis
in the leading and the trailing tails are offset from the progenitor
by different amounts. More important, perhaps, are the selection
biases inherent to the SDSS spectroscopic sample. Finally, the
simplicity of our decomposition of the Sgr MDF into individual
components should not be understated. Nonetheless, we believe that, to
first approximation, the model described below attempts to account for
possible MDF variations caused by the three factors described above, as
it allows the relative contribution of each sub-population to change
from bin to bin in $\tilde{\Lambda}_\odot$

\begin{figure}
\centering
\includegraphics[width=0.4\textwidth]{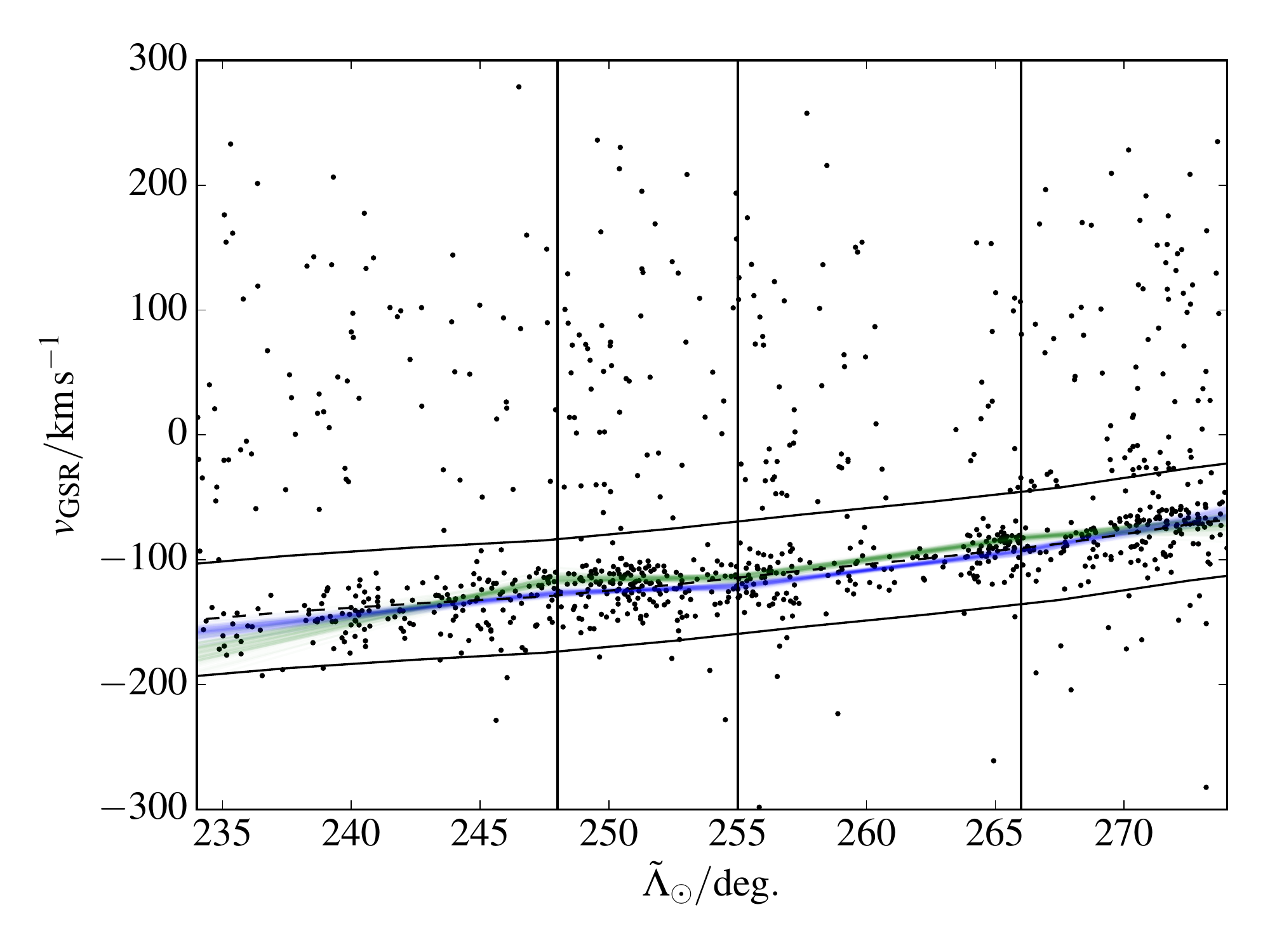}
\includegraphics[width=0.4\textwidth]{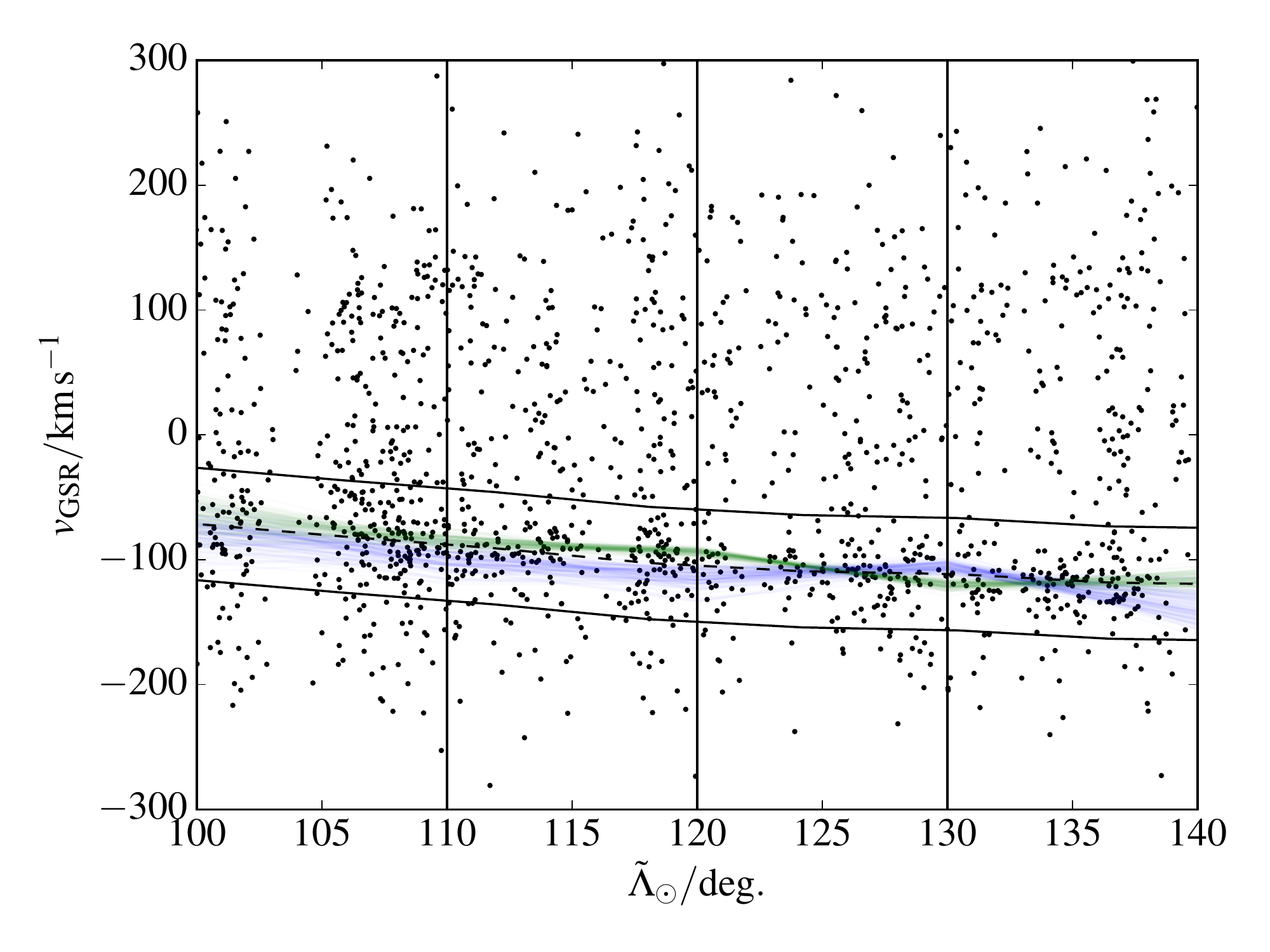}
\caption{Model mean velocity track and bin walls for the trailing
  (upper panel) and leading (lower panel) streams. The solid vertical
  lines indicate the bin walls, the dashed black lines show the
  mean velocity of the track from \citet{Belokurov2014}. The solid
  black lines show the region in velocity space used for selecting
  stream stars to determine the stream's MDF components. Finally, the
  blue and green lines show 100 samples of the mean velocity track
  from the posterior of our chemo-dynamical modelling for the low and
  high metallicity populations.}
\label{fig:vmeans_trailing}
\end{figure}
\begin{figure*}
\centering
\includegraphics[width=\textwidth]{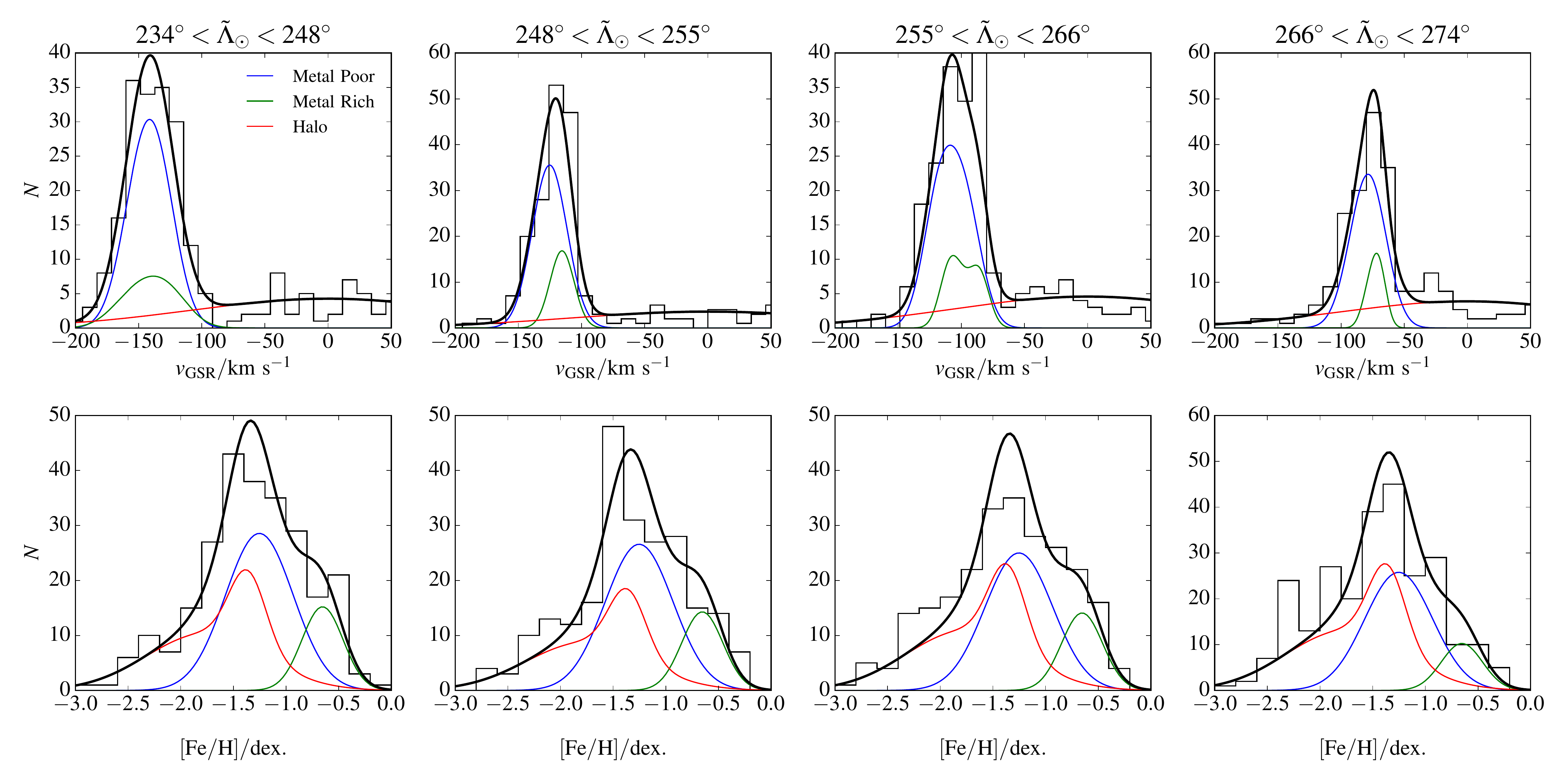}

\caption{The best fitting components of the model in velocity (top
  row) and metallicity space (bottom row) for the trailing stream.
  The blue and green lines shows the inferred distributions of low and
  high metallicity stream components. The red lines display the halo
  component.}
\label{fig:modelmix_trailing}
\end{figure*}
\begin{figure*}
\centering
\includegraphics[width=\textwidth]{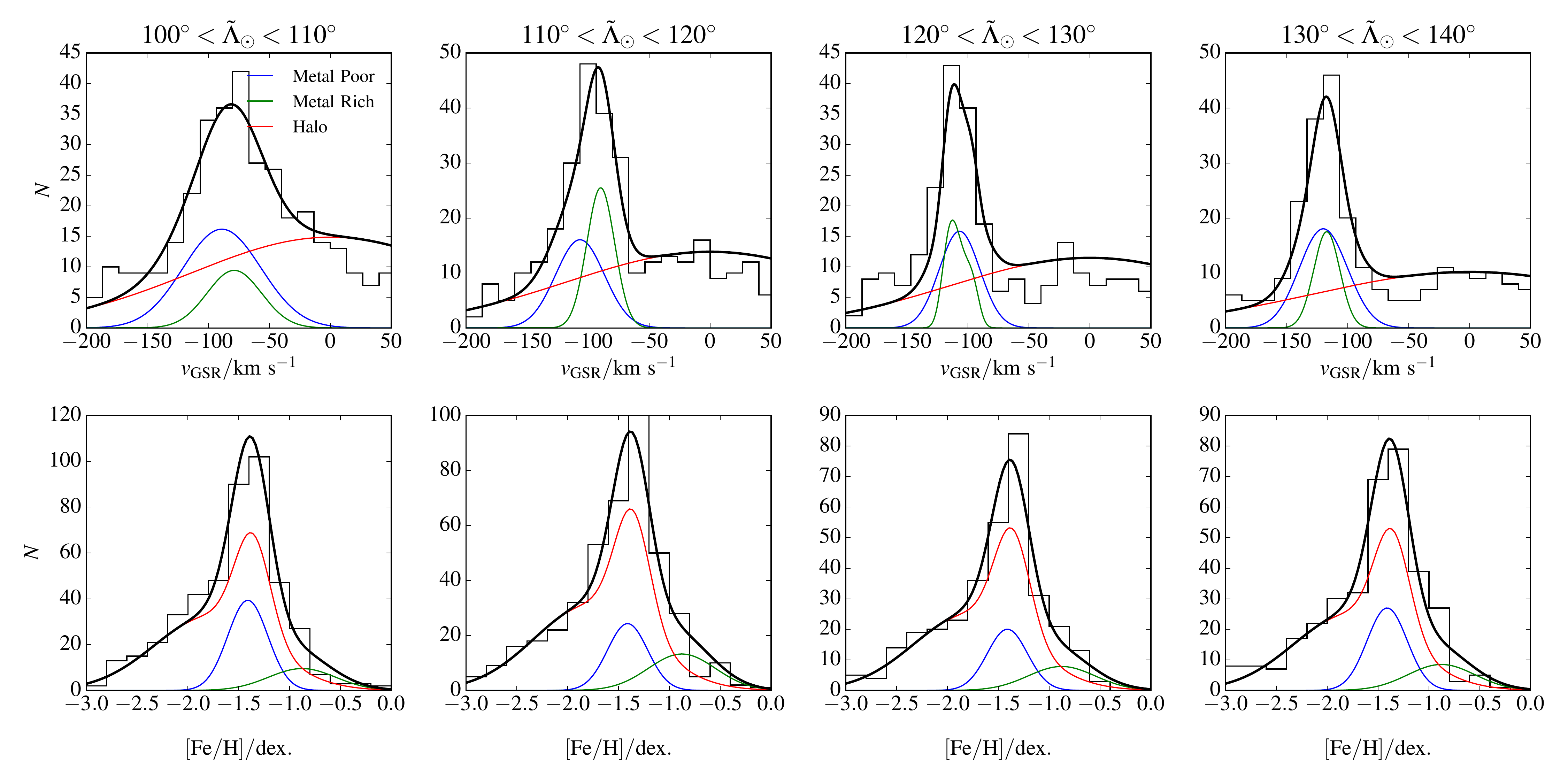}
\caption{As Fig.~\ref{fig:modelmix_trailing}, but for the leading arm.}
\label{fig:modelmix_leading}
\end{figure*}

\begin{figure*}
\centering
\includegraphics[width=\textwidth]{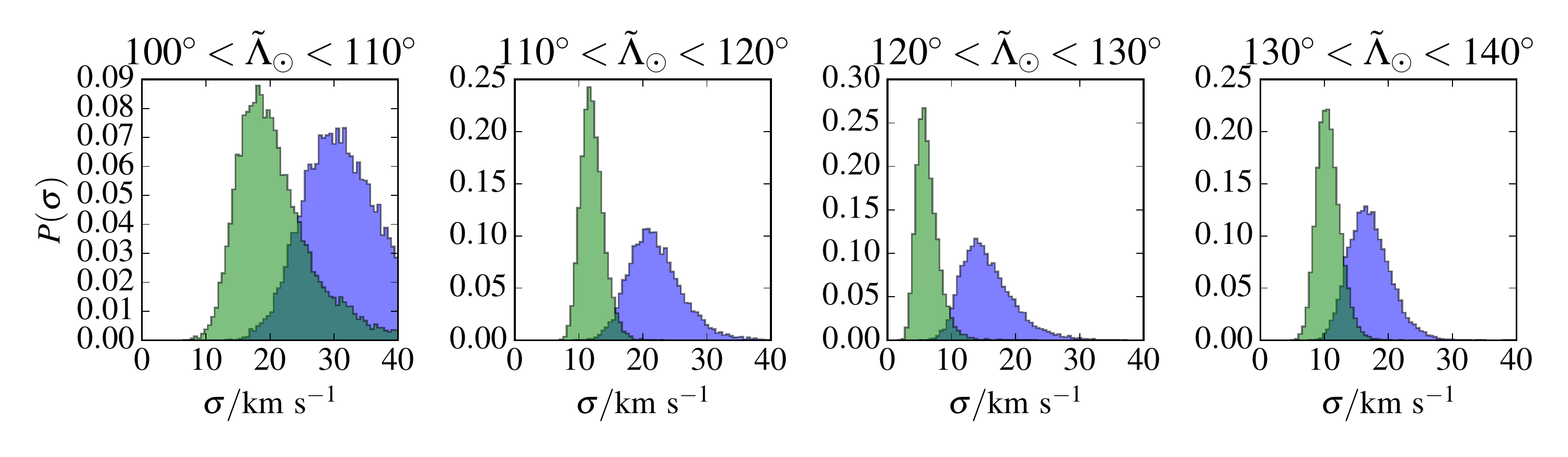}
\includegraphics[width=\textwidth]{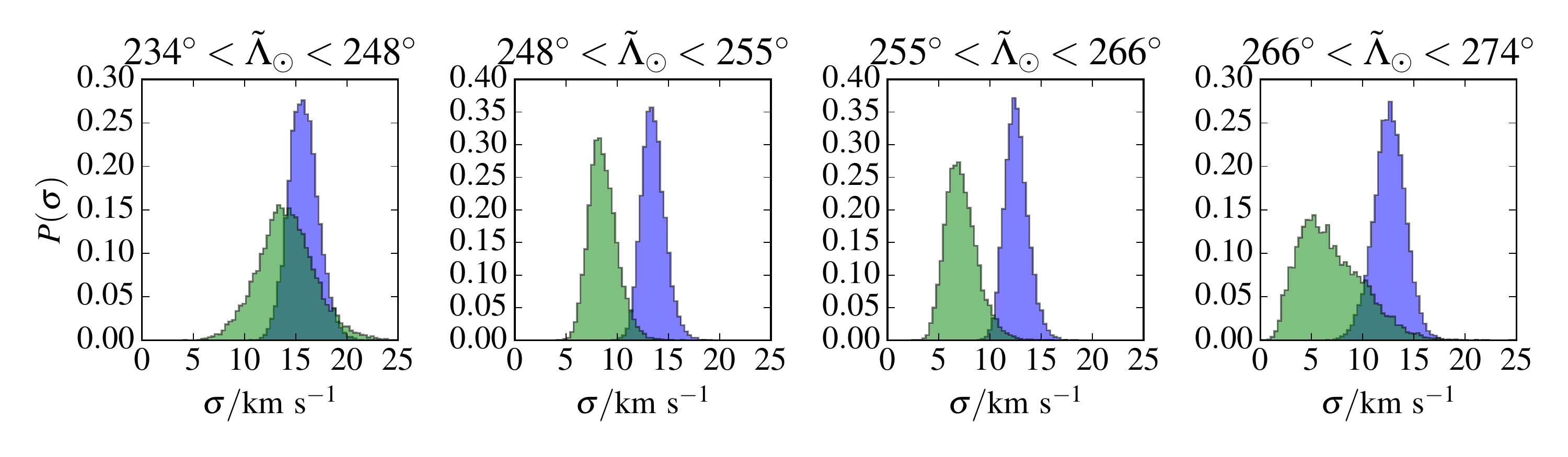}
\caption{The posterior probability distributions for the velocity
  dispersions of the low metallicity (blue) and high metallicity
  (green) components of the leading (upper row) and trailing (bottom
  row) stream. The low metallicity stars show a significantly enhanced
  velocity dispersion, compared with the high metallicity
  stars. Additionally the leading stream shows a systematically higher
  line of sight velocity dispersion compared to the trailing stream.
}
\label{fig:sigma_posterior}
\end{figure*}

\subsection{A Link between Chemistry and Kinematics}

With the metallicity components of the stream in hand, we are now able
to link the stream chemistry with the kinematics. Namely, we strive to
determine the mean velocity and its dispersion for each of the two
metallicity sub-populations along the stream. We model the chemistry
and kinematics in 4 individual bins of $\tilde{\Lambda}_\odot$ for
both trailing and leading tails. We choose the bin boundaries as
follows: $(234^{\circ}, 248^{\circ})$, $(248^{\circ}, 255^{\circ})$,
$(255^{\circ}, 266^{\circ})$, $(266^{\circ}, 274^{\circ})$ for
trailing, and $(100^{\circ}, 110^{\circ})$, $(110^{\circ},
120^{\circ})$, $(120^{\circ}, 130^{\circ})$ and $(130^{\circ},
140^{\circ})$ for leading. Note that the trailing tail bin sizes are
slightly different from one another to ensure that a similar number of
stars are contained in each bin. We model the velocity centroid of
each sub-population of each tail independently. We assume that the
mean velocity track of each component is well-represented by a
piecewise linear function. Within each bin of $\tilde{\Lambda}_\odot$,
we assume that the velocity distribution of the $j$-th sub-population
about the linear trend is Gaussian, with a mean $\mu^{\rm v}_j$ and a
dispersion $\sigma^{\rm v}_j$. Additionally, we assume that the
contaminant population has a zero mean velocity and a dispersion of
$\sigma^{\rm v}_{\rm MW}$ km s$^{-1}$. Thus, within each bin of
$\tilde{\Lambda}_\odot$, the likelihood of observing the $i$-th star
with velocity $v_{\rm GSR,i}$ and metallicity [Fe/H]$_i$ is simply:
\begin{equation}
\mathcal{L}_i = \sum_{j=1}^3 \frac{w_j P_j([{\rm Fe/H}]_i)}{\sqrt{2\pi
    [(\sigma^{\rm v}_j)^2 + (\sigma^{\rm v}_i)^2]}} \exp \left(
-\frac{[v_{{\rm GSR}, i} - \mu^{\rm v}_j(\tilde{\Lambda}_\odot)]^2}{2[(\sigma^{\rm
      v}_j)^2 + (\sigma^{\rm v}_i)^2]} \right)
\end{equation}
Here, $\sigma^{\rm v}_i$ is the radial velocity error of $i$-th star,
$P_j([{\rm Fe/H}]_i)$ is the probability of observing a star with
metallicity $[{\rm Fe/H}]_i$ belonging to the $j$-th
sub-population. This probability is Gaussian whose position and shape
is fixed by the coefficients determined in the previous
sub-section. To clarify, the index $j$ runs from 1 to 3, corresponding
to the two stream and one foreground components.  More specifically,
$j=1$ labels the metal-poor, $j=2$ the metal-rich sub-populations,
while $j=3$ labels the MW. We enforce that sum of the relative weights
equals to 1, i.e. $w_j=1$. In total, for each tail we have 30
independent free parameters encompassing 4 bins in
$\tilde{\Lambda}_\odot$: 8 weights, 12 velocity dispersions and 10
mean stream velocities at 5 bin walls. All of the kinematic parameters
are constrained simultaneously. The results of the fit are presented
in Tables~\ref{tab:model_trailing} and \ref{tab:model_leading} for the
trailing and leading tails respectively.

Fig.~\ref{fig:vmeans_trailing} shows the behaviour of the mean of the
velocity of the trailing and leading tails correspondingly as a
function of $\tilde{\Lambda}_\odot$. The model prediction for the
stream's metal-poor component is shown in blue, and for the metal-rich
in green. The differences between the velocity tracks of the two
sub-populations are typically rather small, of order of $10$ km
s$^{-1}$. Note, however, that even such a small velocity mismatch
could be enough to inflate the velocity dispersion if it was assumed
to be the same for all stream members indiscriminately.

\begin{figure}
\centering
\includegraphics[width=0.49\textwidth]{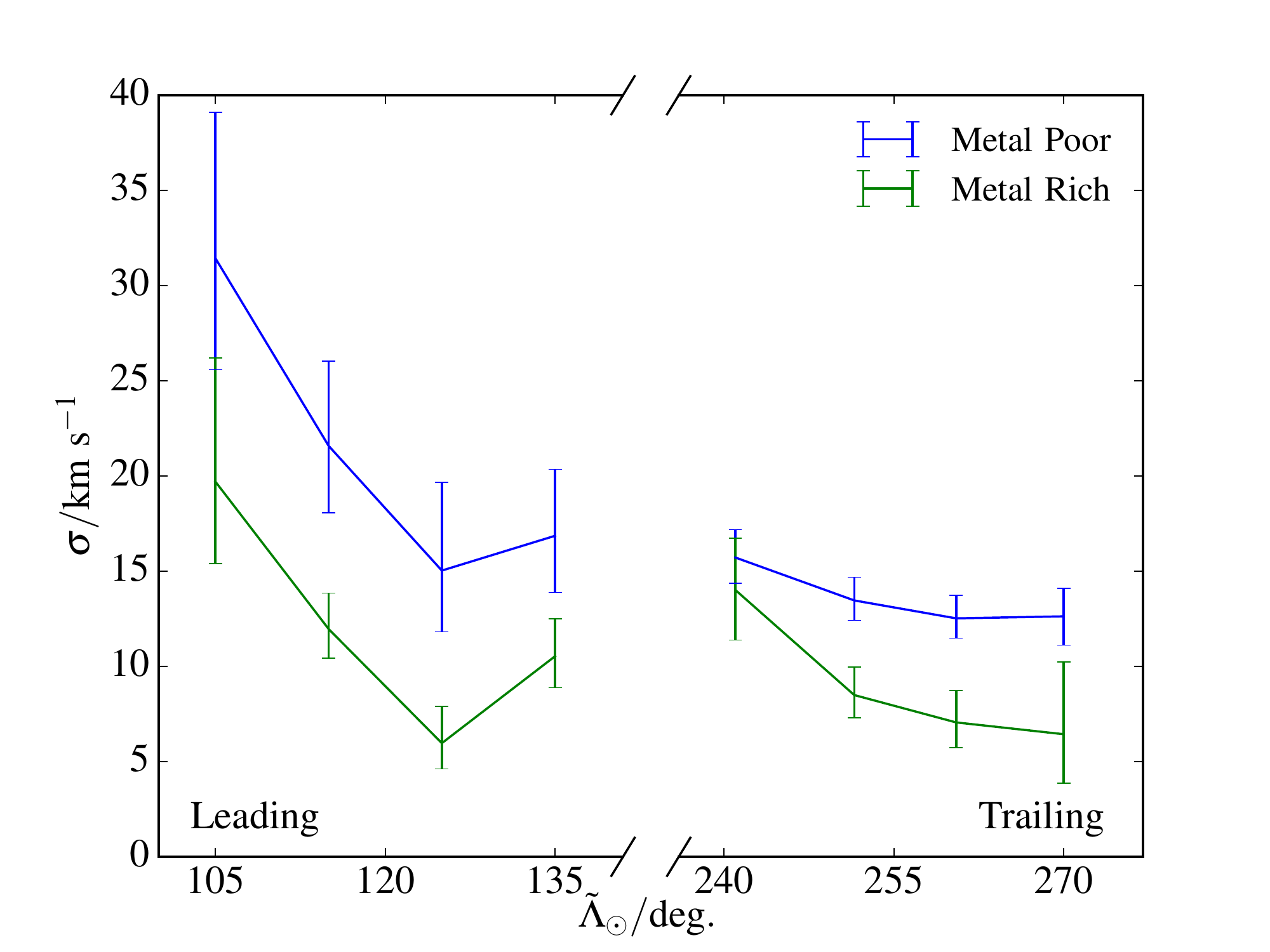}
\caption{Variation of the velocity dispersions in the metal-rich and
  metal-poor components as a function of the Sgr stream longitude for
  both leading and trailing stream.}
\label{fig:sigma_lambda}
\end{figure}

Figs~\ref{fig:modelmix_trailing} and \ref{fig:modelmix_leading} give a
detailed view of the velocity (top row) and the metallicity (bottom
row) distributions in each $\tilde{\Lambda}_\odot$ bin in the trailing
and leading tails correspondingly. As before, black histograms are the
data, and the blue (green) curves show models for the metal-poor
(metal-rich) sub-populations. Finally, the red line displays our model
for the MW halo contribution. As is evident from the Figures, the
observed velocity and metallicity distributions are reproduced
adequately. This is helped by the fact that both in the North and the
South, the Sgr stream stands out kinematically from the bulk of the
Galactic foreground. While the gross features of the distributions are
never amiss, some of the finer details are clearly not captured by our
model. This is hardly surprising though, as we have postulated that
the MW halo can be represented with only one Gaussian component in
velocity (although its width is allowed to vary from bin to bin), and
a combination of two broad Gaussians in metallicity. However, it is
likely that there exists unmixed - and thus probably highly peaked in
terms of their MDF and phase-space distribution - stellar halo
sub-structures in the areas of the sky under consideration. Generally,
these would include portions of unrelaxed stellar debris similar to
those detected by \citet[][]{echos1, starkenburg2009,
  DeasonHalo,torrealba2015, Janesh2016}. Additionally, there exist
large known structures overlapping with the leading tail in the North,
such as the Virgo Cloud \citep[see
  e.g.][]{Duffau2006,Newberg2007,Martinez2007,Juric2008}; and the
trailing tail in the South, such as the Cetus stream
\citep[][]{Newberg2009,SGB-trailing-paper}.

According to Fig.~\ref{fig:modelmix_trailing} and
Table~\ref{tab:model_trailing}, the trailing tail data in the South
suffers very little MW foreground contamination. The stream runs
across the four $\tilde{\Lambda}_\odot$ bins with substantial negative
velocity. Typically, in each bin approximately half of the stars in
our final sample belongs to the Sgr stream with $\sim70\%$ in the
metal-poor sub-population and $\sim30\%$ in metal-rich. The other half
of the sample resides in the MW component, for which we estimate the
velocity dispersion to be in the range $100$kms$^{-1}<\sigma^{\rm
  v}_{\rm MW} <115$ kms$^{-1}$. A striking result of the modelling is
the substantial difference between the kinematical properties of the
two Sgr stream sub-populations split by metallicity. As mentioned
before, there exists a $\sim 10$ km s$^{-1}$ offset between their mean
velocities. More intriguingly, the sub-populations appear to have
rather different velocity dispersions. The metal-rich sub-population
exhibits kinematics well approximated by a Gaussian with
$\sigma^{\rm v}_{\rm ST2} \sim 8$ km s$^{-1}$. This is in good
agreement with the value reported by \citet[][]{Monaco2007} using high
resolution spectroscopy of (mostly) M-giant stars in the neighbouring
part of the sky. The metal-poor component is almost twice as hot, with
$\sigma^{\rm v}_{\rm ST2} \sim 13$ km s$^{-1}$, in consensus with the
measurements of \citet[][]{Koposov2013} who also used the SDSS spectra
in their analysis, albeit without distinguishing between the two
metallicity populations.

Fig.~\ref{fig:modelmix_leading} and Table~\ref{tab:model_leading}
present the results of the leading tail modeling. In the North, the
Sgr stream constitutes only $25\%$ of the (cleaned) spectroscopic
sample, with the two metallicity sub-populations contributing almost
equal amounts: $\sim60\%$ is metal-poor and $\sim40\%$ is
metal-rich. Note, however, that as described above, the actual
definitions of the ``metal-rich'' and ``metal-poor'' are now different
compared to the trailing tail data. Interestingly, we find a slightly
hotter MW halo, with the velocity dispersion in the range $110$
kms$^{-1}< \sigma^{\rm v}_{\rm MW} < 130 $kms$^{-1}$. The dispersion
of the Sgr stream itself is also inflated: we find $15$ kms$^{-1}<
\sigma^{\rm v}_{\rm SL1} < 30 $kms$^{-1}$ for the metal-poor
sub-population, and $6$ kms$^{-1}< \sigma^{\rm v}_{\rm SL2} < 20
$kms$^{-1}$ for the metal-rich.

Fig.~\ref{fig:sigma_posterior} displays the posterior distributions
for the Sgr stream velocity dispersion in both leading and trailing
tails in each bin of $\tilde{\Lambda}_\odot$ studied. In all bins and
in both tails, the picture remains unchanged: the metal-poor
sub-population exhibits an enhanced velocity dispersion compared to
the metal-rich one. While in all cases there is a noticeable overlap
between the posterior distributions, their peaks are typically
$2\sigma$ apart. Figure~\ref{fig:sigma_lambda} summarizes the velocity
dispersion behaviour as a function of $\tilde{\Lambda}_\odot$. There
appears to be a clear evolution of the dispersion with the stream
longitude in both the leading and the trailing tails: the dispersion
increases towards lower $\tilde{\Lambda}_\odot$.

Some of the change in the line-of-sight velocity dispersion of the
debris along the stream is likely due to the projection effects. For
different $\tilde{\Lambda}_\odot$, the line of sight from the
heliocentric observer pierces the stream at different
angles. Therefore, the dispersion is expected to be higher for
directions where the angle between the line of sight and the debris
orbital velocity is smaller. For the trailing tail, this occurs around
$\tilde{\Lambda}_\odot \sim 240^{\circ}$. Additionally, the intrinsic
velocity dispersion of the stars in the stream is expected to grow due
to the debris evolution in the gravitational potential. In the
simplest case, in a spherical potential, the stream's cross-section
and velocity dispersion in the radial direction will tend to inflate
due to differential apsidal precession \citep[see
  e.g.][]{Johnston2001,Hendel2015}.  Moreover, as shown in simulations
of the Sgr disruption \citep[see e.g.][]{LM2010,Gibbons2014}, the
stream separates into individual streamlets, consisting of the debris
unbound during subsequent pericentric passages \cite[see
  also][]{nicola_streams}. With time, these ``feathers'' will misalign
-- also due to the differential apsidal precession -- making the stream
appear ruffled around the apogalacticon, and thus contributing to the
line-of-sight velocity dispersion increase. As measured by
\citet[][]{Belokurov2014}, the leading tail apocenter lies around
$\tilde{\Lambda}_\odot \sim 70^{\circ}$, hence some of the dispersion
rump-up around $\tilde{\Lambda}_\odot \sim 100^{\circ}$ could be due
to the superposition of stars stripped at different epochs. Finally,
in a flattened potential, the differential orbital plane precession
will lead to the debris fanning in the direction perpendicular to the
stream plane, therefore also inflating the line-of-sight velocity
dispersion \citep[see e.g.][]{Erkal2016}. As pointed out by these
authors, the amount of debris fanning varies strongly with the angle
along the stream.


%
\begin{figure*}
\centering
\includegraphics[width=0.9\textwidth]{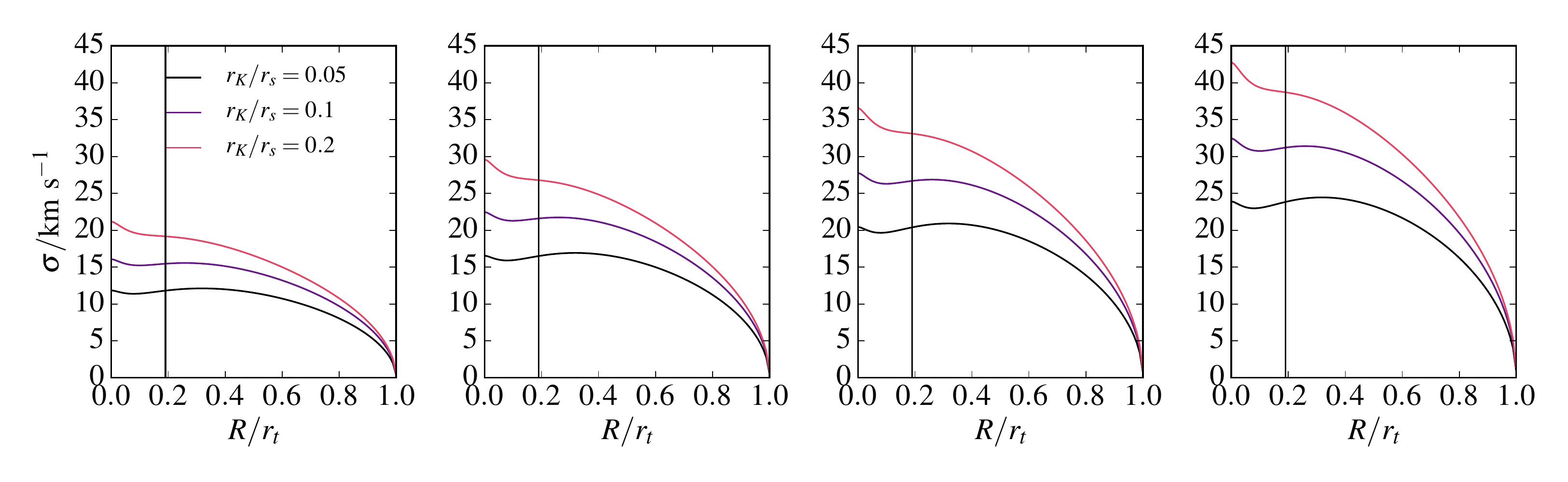}\\
\includegraphics[width=0.9\textwidth]{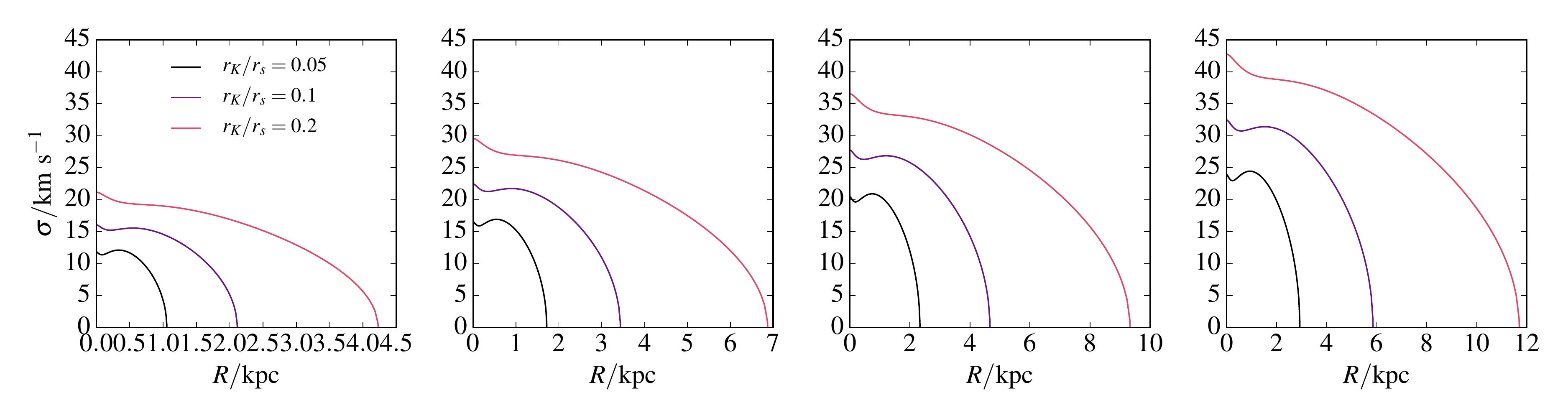}\\
\includegraphics[width=0.9\textwidth]{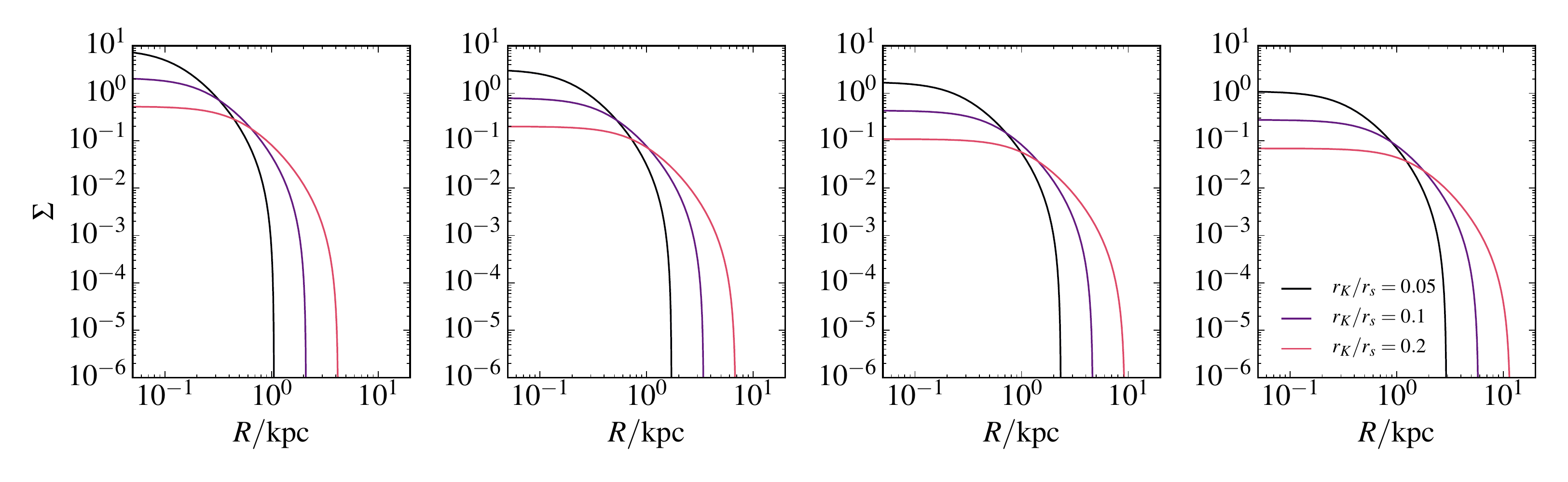}
\caption{{\it Top and Middle:} the initial projected velocity
  dispersions for each of the progenitor models, with the top row
  showing distances in units of tidal radius $r_t$ and the middle row
  in kpc. The panels from left to right show the four progenitor
  masses that we consider, $1\times10^{10} M_\odot$ to $1\times
  10^{11} M_\odot $ and each of the coloured lines correspond to the
  degree of embeddedness ($\rK / \rs$) of the stellar particles within
  each halo. The vertical black line indicates the half light
  radius. Note that the King model embedded within an NFW halo
  produces velocity dispersion profiles that are flat out to almost
  the tidal radius. {\it Bottom:} initial surface brightness profiles
  in the simulated Sgr dwarf progenitors.  The overall range of
  half-light radii probed by the 12 models is from 0.2 kpc to 2.2
  kpc.}
\label{fig:sigma_profiles}
\end{figure*}
\begin{figure*}
\centering
\includegraphics[width=0.49\textwidth]{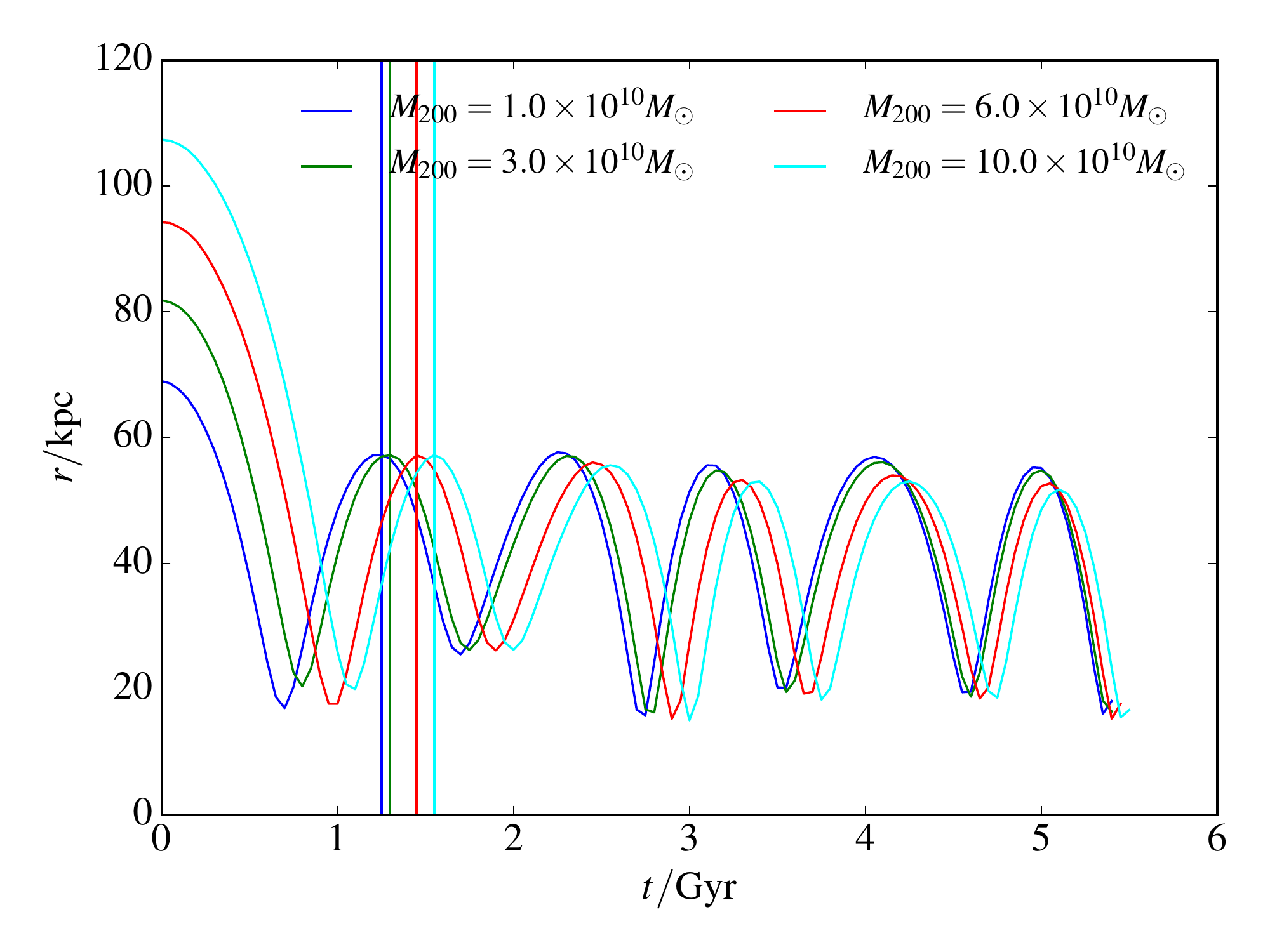}
\includegraphics[width=0.49\textwidth]{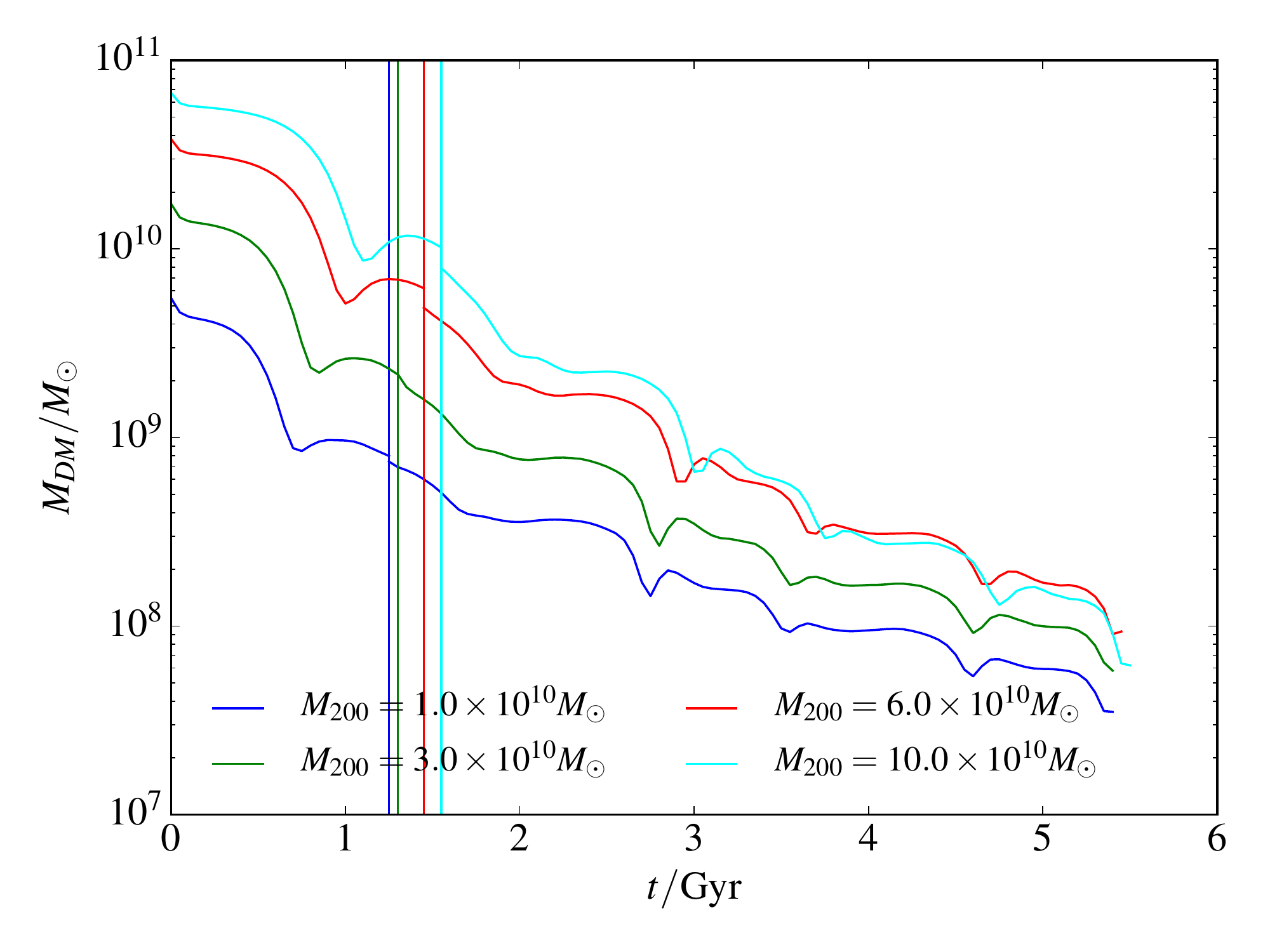}
\caption{The orbit of Sgr (left) and the evolution of the DM mass
  within the tidal radius (right). The more massive the progenitor,
  the more effective dynamical friction is at reducing its orbit. The
  more massive the progenitor, the more ruthless the stripping so that
  all four models yield a present-day remnant with dark matter mass
  $\lesssim 10^8 M_\odot$.}
\label{fig:orbitSgr}
\end{figure*}
\begin{figure*}
\centering
\includegraphics[width=0.99\textwidth]{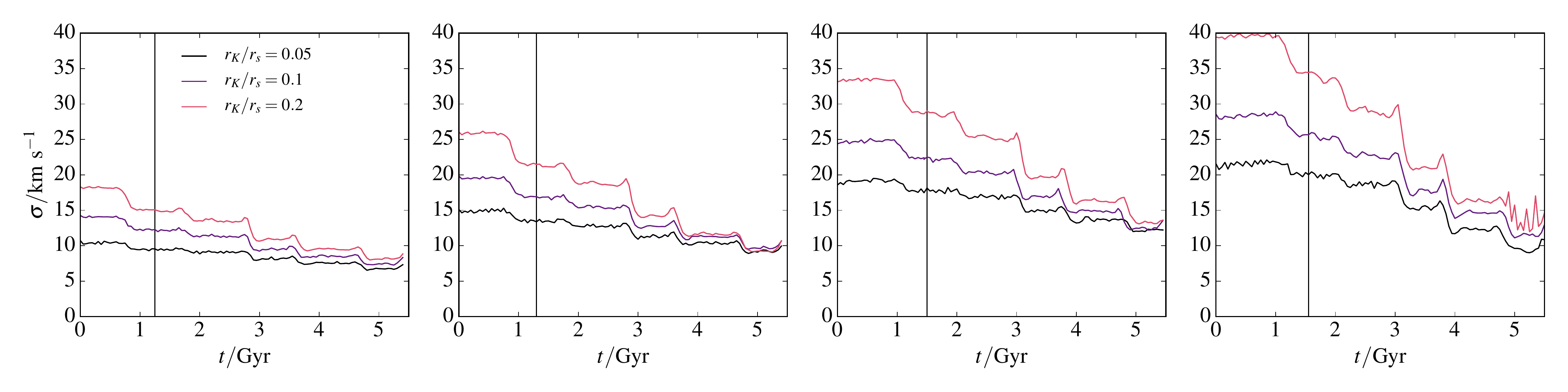}\\
\includegraphics[width=0.99\textwidth]{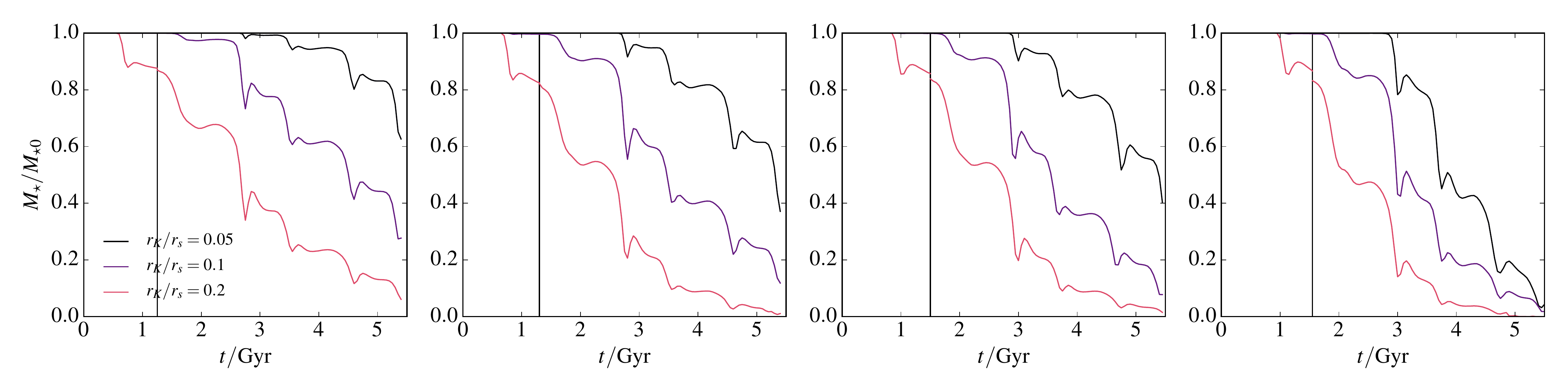}
\caption{Evolution of the velocity dispersion and stellar mass. The
  vertical black line indicates the crossover from the TF potential to
  the \citet{LM2010} potential.  The panels from left to right show
  the four progenitor masses that we consider, from $1\times10^{10}
  M_\odot$ to $1\times 10^{11} M_\odot $. The coloured lines show
  different embeddedness $\rK/\rs$ of the stellar distribution. Note
  that based on the current observations of the stream and the
  remnant, the Sgr is estimated to have lost $\sim50\%$ of its stellar
  mass. While this could easily be the upper bound, we point out that
  some of our models have ended up losing more stars that have been
  observed so far.}
\label{fig:sim_properties}
\end{figure*}

\section{From Dwarf to Stream}
\label{sec:sims}

Clearly, the debris velocity dispersion of $\sim15$ kms$^{-1}$ must
imply a progenitor more massive than $\sim 10^9$ M$_{\odot}$ as
inferred from the measurement of 8 kms$^{-1}$. However, the link
between the progenitor's original mass and the final velocity
dispersion in the stream is quite labyrinthine. First, rather than the
dispersion of the stripped DM particles, we are interested in the
kinematics of the {\it stellar} debris, whose properties will depend
on how deeply embedded the progenitor's stellar component is within
the parent DM halo. Second, the progenitor's velocity dispersion will
evolve during the disruption due to mass loss, which will be reflected
in the kinematics of the tails. Importantly, the mass loss rate
depends on the orbit of the progenitor, which will be affected by
dynamical friction. Finally, the direct comparison to the data is only
possible if the stream's 3D geometry matches that of the observed
stream -- otherwise projection effects might bias the inference.

In an attempt to address all of the factors mentioned above, we have
settled on the following setup for the Sgr accretion simulation. We
explore the behaviour of streams produced through tidal stripping of
progenitors of different masses. For each mass, we use three different
(massless) stellar components with different sizes. We start our
simulations in live DM halos to facilitate accurate modelling of
dynamical friction. Once the bulk of the mass is tidally stripped, and
the orbits have largely finished their evolution, we move the
remaining bound particles into the \citet{LM2010} (parameterized)
potential. The last step ensures that the geometry of the simulated
tails is correct and we do not need to worry about the projection
effects. 

\subsection{The Progenitor Model}

In our model, the dark halo of the Sgr progenitor follows a NFW
\citep{NFW1996} profile with a density given by
\begin{equation}
\rho_{\rm NFW} = \frac{M_{200}}{4 \pi \rs^2} \frac{(r/\rs)^{-1} (1 +
  r/\rs)^{-2}}{\log(1+c_{200}) - c_{200}/(1+c_{200})}
\label{eq:nfwdens}
\end{equation}
where $\rs$ is a scale radius, $M_{200}$ is the mass contained within
a radius, $r_{200}$ where the density of the halo is 200 times the
critical density of the universe and $c \equiv r_{200} / \rs$ is the
NFW concentration parameter.  To reduce the parameter space through
which we need to search, and to ensure that the parameters of our
progenitor's halos are well motivated cosmologically, we link the
masses and concentrations of the progenitors halo through the
mass-concentration relation of \citet{Maccio2007}. The NFW profile of
eq.~(\ref{eq:nfwdens}) has an infinite mass. To allow us to set up
equilibrium realisations, we exponentially truncate the dark halo's
density at the virial radius following the prescription of
\citet{Kazantzidis2004}.

The stellar component follows a \citet{King1962} profile, with mass
density
\begin{equation}
\rho_{\rm K} = \frac{K}{x^2} \left[\frac{\arccos(x)}{x} - \sqrt{1-x^2} \right] ~~ ;~~
x \equiv \left[\frac{1 + (r/\rK)^2}{1 + (\rt / \rK)^2}\right]^{1/2}
\label{eq:kingdens}
\end{equation}
where $\rt$ is the tidal radius, $\rK$ is a scale radius and $K$ is an
arbitrary constant. In all of our simulations, we choose to keep the
ratio $\rK / \rt$ to be fixed at 8, in keeping with the fits of
\citet{Majewski2003} to the light profile of the Sgr dwarf at the
present day.  We vary the degree of `embeddedness' of the light
profile in the dwarf, described by the ratio $\rK / \rs$, choosing
values in the range $0.05 - 0.2$. These choices yield projected
velocity dispersion profiles for the stellar component that are flat
out to a few half-light radii. This is shown in the upper panels of
Fig.~\ref{fig:sigma_profiles}, in which the half-light radius is
marked by a vertical line.  If the Sgr progenitor was a dSph, this is
a natural assumption to make. The pioneering work of \citet{Kl02} on
Draco provided the first evidence of extended flat velocity dispersion
profiles, and this was confirmed by the extensive study of
\citet{Wa07}. Recent modelling of dwarf spheroidals also lays emphasis
on the very flat dispersion profiles within the radii probed by
existing observations~\citep{Am11,Bu15}. Hence, as
Fig.~\ref{fig:sigma_profiles} confirms, our progenitor looks very much
like a dSph in terms of both its photometric and kinematic properties.

The velocity distribution for the luminous King component and the dark
halo is obtained assuming isotropy and using the Eddington's
formula~\citep{GalacticDynamics}, namely
\begin{equation}
f_i(\epsilon) = \frac{1}{\sqrt{8} {\rm \pi}^2} \int_0^\epsilon
\frac{\dd^2 \rho_i}{\dd \Psi^2} \frac{\dd \Psi}{\sqrt{\epsilon -
    \Psi}}
\label{eq:eddington}
\end{equation}
Here, $\epsilon$ is the binding energy and $\Psi$ the potential, which
is taken as that of the NFW dark halo only, as it dominates the
luminous compoment. Here, $i$ refers to either the King or the NFW
densities in eqs.~(\ref{eq:nfwdens}) or ~(\ref{eq:kingdens}).

\subsection{The Milky Way Models}

The Milky Way is modelled in two ways. For the final 4 Gyr of its
orbit, the disrupting Sgr is evolved in the \citet{LM2010}
potential. Although the details of this potential may not be correct,
it has been shown to provide a good match to much of the data on the
Sgr stream. The potential is conventional in its choice of disk and
bulge, using a Miyamoto \& Nagai disk and Hernquist sphere
respectively. Unusually, though, the dark halo is misaligned, triaxial
and logarithmic of form
\begin{equation}
\Psi = - v_0^2 \log \left( C_1 x^2 + C_2 y^2 + C_3 xy + z^2/q^2 + r_{\rm
  halo}^2 \right ).
\end{equation}
The choice of parameters $C_i, q$ and $r_{\rm halo}$ yields a triaxial
dark matter halo whose minor to major axis ratio $(c/a)_\Phi = 0.72$
and intermediate to major axis ratio $(b/a)_\Phi = 0.99$ at radii 20
kpc$ < r < $ 60 kpc, with the minor, intermediate and major axes of
this halo lie along the directions ($\ell, b$) = ($7^\circ, 0^\circ$),
($0^\circ, 90^\circ$) and ($97^\circ, 0^\circ$) respectively.

For the initial phase of the simulation, where the Sgr dwarf falls
into the Milky Way, we represent the potential using a live
realisation of the ``truncated flat rotation curve'' (TF) models
presented in \citet{Gibbons2014}. These are a three parameter family
of models representing the total matter content of the galaxy, with a
rotation curve that is flat, with amplitude $v_0$ in their inner part,
which then smoothly transitions at a scale radius $\rTF$ into a power
law decline in their outer parts with slope $\alpha$.  The rotation
curves of these models have the form
\begin{equation}
v_c^2 = \frac{v_0^2 \rTF^\alpha}{\left(\rTF^2 + r^2\right)^{\alpha/2}}
\end{equation}
which implies a density
\begin{equation}
\rho_{\rm TF} = \frac{v_0^2}{4\pi G} \frac{1 - (\alpha -
  1)(r/\rTF)^2}{r^2 [(r/\rTF)^2 + 1]^{\alpha / 2 + 1}}
\end{equation}
We choose the parameters $v_0=230 \kms$, $\rTF=15$ kpc and $\alpha=0.5$
corresponding to the most likely solution from \citet{Gibbons2014}.

In this form, the TF model has a divergent mass for $\alpha < 1$. To
be able to construct an equilibrium realization of this density, we
must truncate it. We do this by exponentially truncating the density
at a radius $\rtr$, with a decay scale of $\rd$, whilst ensuring that
the resulting density has a continuous logarithmic derivative. The
modified density is then
\begin{equation}
\rho_{\rm TF, trunc} = 
  \begin{cases}
    \rho_{\rm TF}(r) & r < \rtr,\\
    \rho_{\rm TF}(r) \left(\frac{r}{\rtr}\right)^{\rtr / \rd} \exp
    \left(-\frac{r - \rtr}{\rd}\right)&  r >\rtr.
  \end{cases} 
\end{equation}
To ensure that this truncation procedure has a negligible effect on
the velocity dispersion of the resulting halo, we choose the
trunction radius to be much larger than the initial apocentre of the
largest Sgr orbit that we consider, setting $\rtr = 500 \kpc$ and $\rd
= 15 \kpc$.  The distribution function for this model is then found by
using the Eddington formula (\ref{eq:eddington}).  Following the
example of \citet{Kazantzidis2004}, the $\dd^2 \rho /\dd \psi^2$ term
can be evaluated easily in terms of derivatives of the density with
respect to $r$ (which are available analytically), and expression
involving $\dd \psi / \dd r$ and $\dd^2 \psi / \dd r^2$. These terms
can be expressed in terms of the enclosed mass $M(r)$ and the
density. Thus, the integral in Eddington's formula reduces to a simple
numerical quadrature.

\begin{figure*}
\centering
\includegraphics[width=0.49\textwidth]{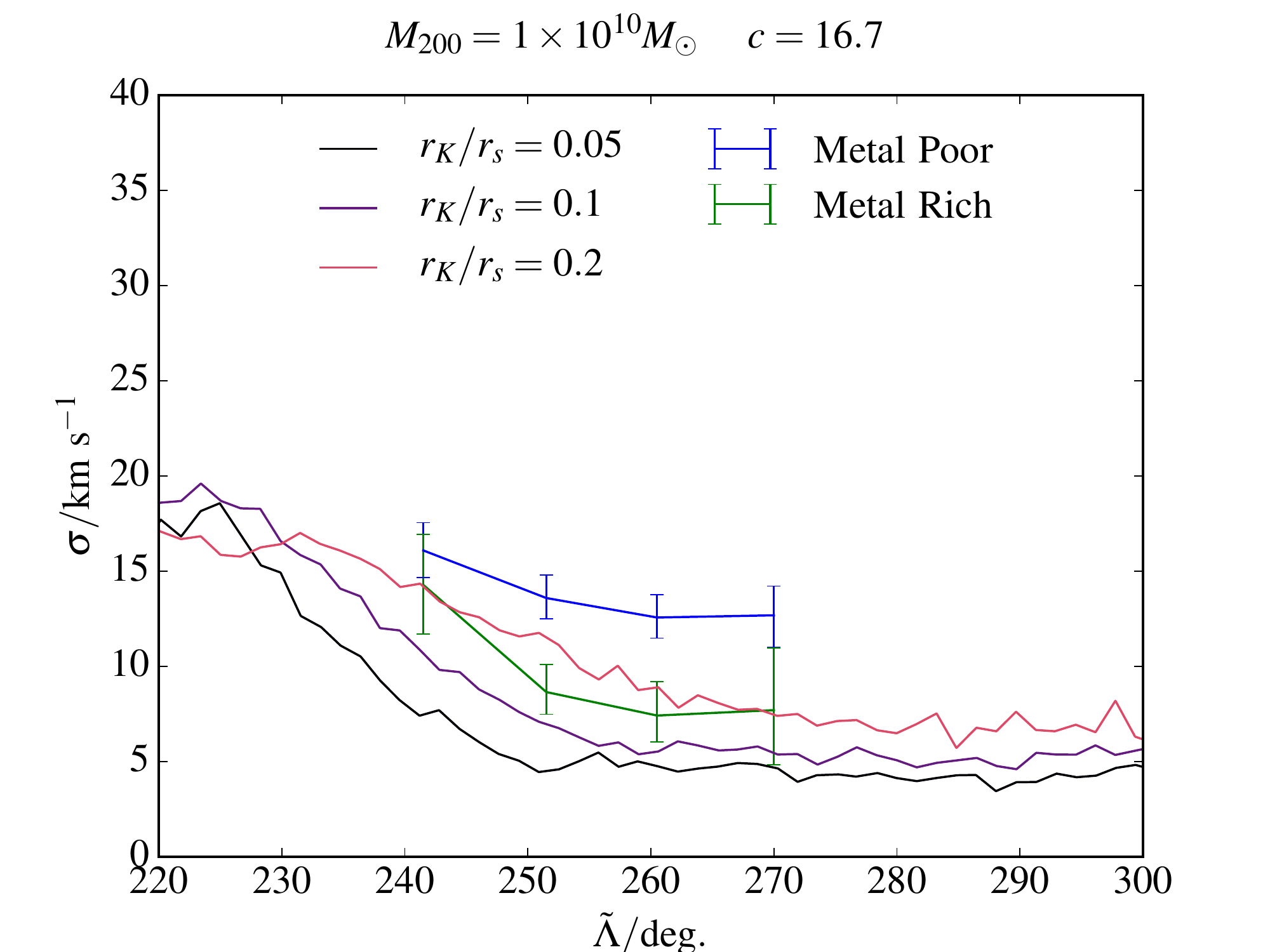}
\includegraphics[width=0.49\textwidth]{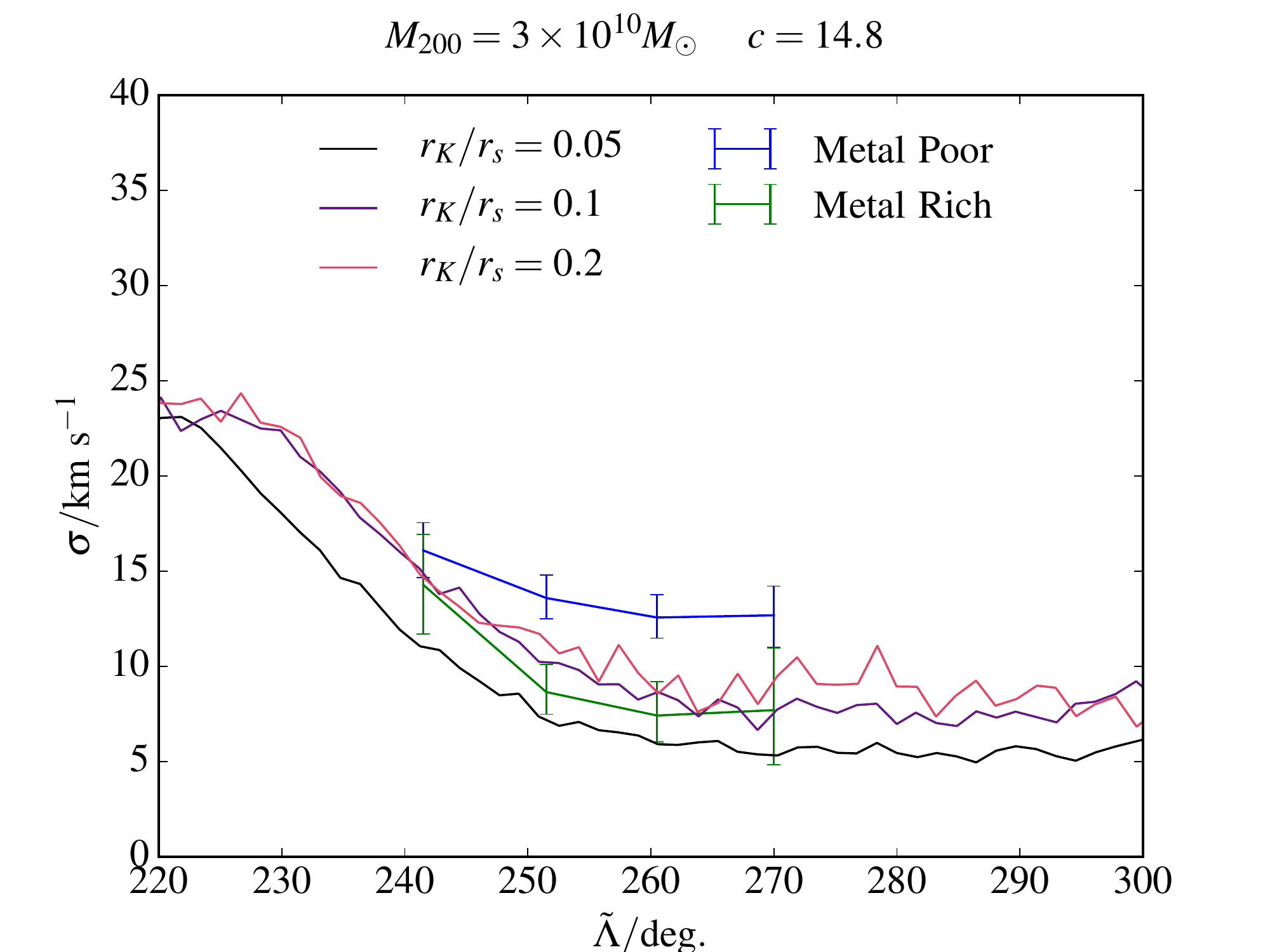}\\
\includegraphics[width=0.49\textwidth]{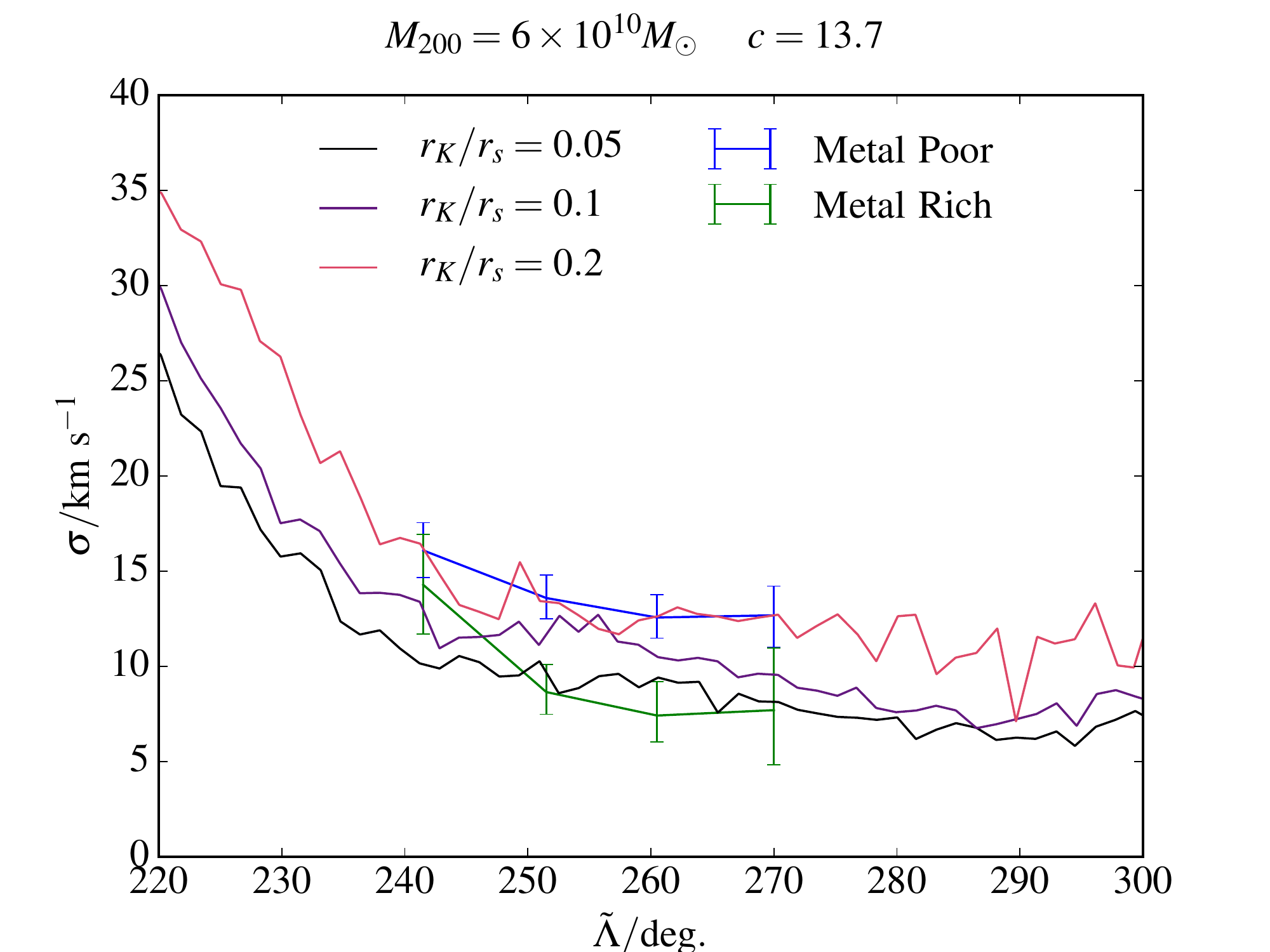}
\includegraphics[width=0.49\textwidth]{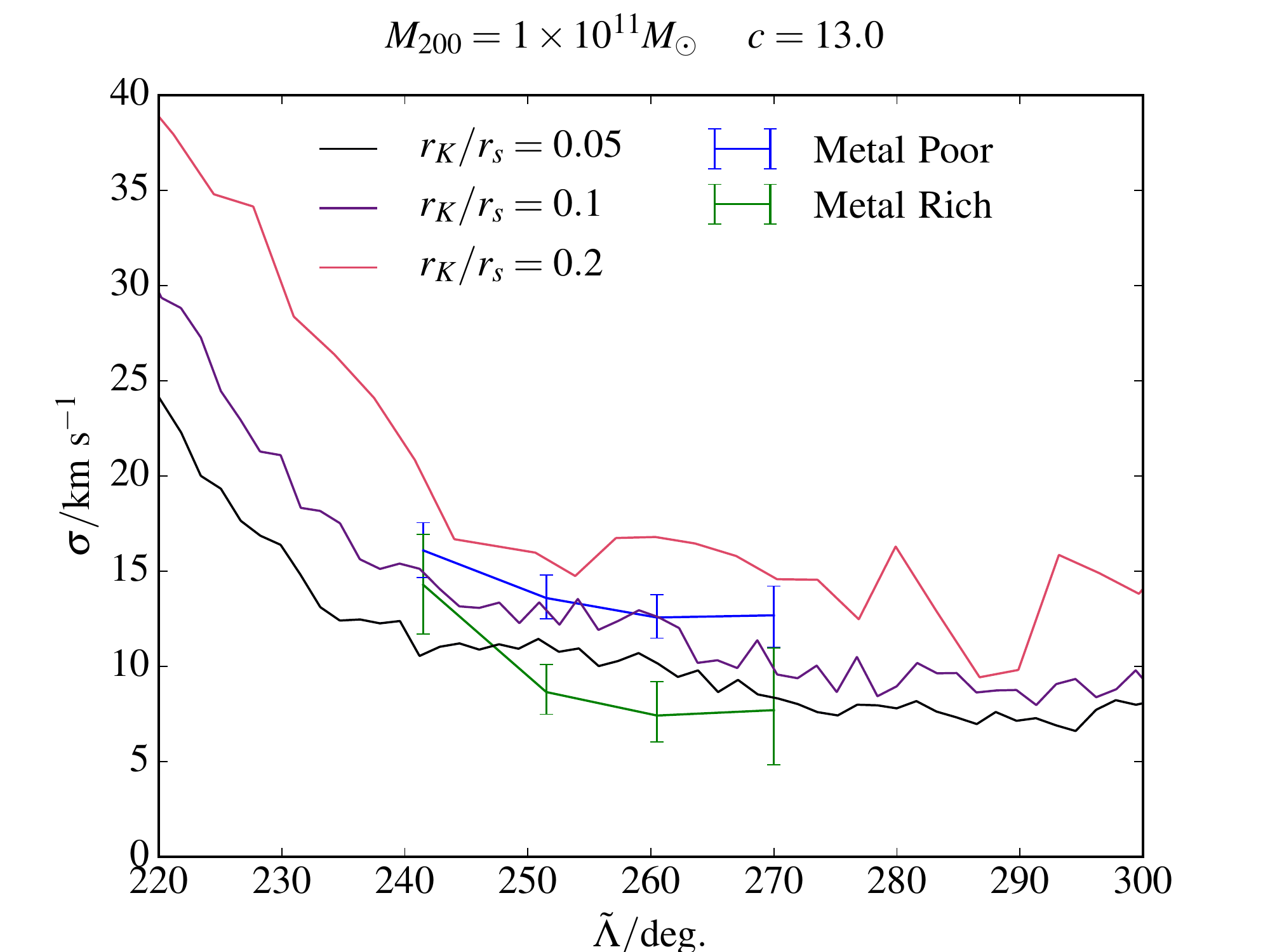}
\caption{Velocity dispersion as a function of longitude in the
  trailing arm. Above each panel is marked the dark halo mass and
  concentration of the Sgr progenitor. The coloured lines show
  different ratios of King radius $\rK$ to dark halo scalelength
  $\rs$, whilst the velocity dispersion data for the metal-poor and
  metal rich stars (measured in this work) are shown in blue and green
  respectively.}
\label{fig:trailvdisps}
\end{figure*}

\subsection{Evolution of the 2-component Sgr model}

The simulation proceeds in 2 phases. First, we evolve the Sgr in the
TF potential until a Crossover Point where most of its DM halo is
stripped and the stellar components begins to disrupt.  By the time
the Sgr gets to the Crossover Point, it has lost between 90\% and 99\%
of its dark matter mass. Most of the orbital evolution happens during
this time.  At the Crossover Point, we take all the bound particles,
(i.e. particles within the tidal radius of Sgr) and place them into
the \citet{LM2010} potential and evolve for 4 Gyr. At the end of the
simulation, the progenitor is near the current Sgr location and the
tidal tails look as they should.  The location of this Crossover
Point is fixed by integrating the orbit of Sgr from its current
position back by 4 Gyr. At this point, Sgr is at
$(X,Y,Z)=(-48,-22,-21)$ kpc and $(V_X,V_Y, V_Z)=(39,-6,-83) \kms$
where $(X,Y,Z)$ is a right-handed Cartesian set with the $X$ axis
pointing towards the Sun.

To match at the Crossover Point, we proceeed as follows. We integrate
the Sgr back in time from Crossover for 1 orbital period with an
approximate prescription for the dynamical friction.  \citet{Ch60}
provided such a formula under the idealized assumption of an
undisruptable stellar system moving through a homogeneous background
of stars with a Maxwellian velocity distribution, namely
\begin{equation}
\frac{\mathrm{d}\mathbf{v}}{\mathrm{d}t} = - \frac{4 \pi G^2 M \rho \ln \Lambda}{v^2} \left[ \mathrm{erf}(X) - \frac{2X}{\sqrt{\pi}} e^{-X^2}\right] \frac{\mathbf{v}}{v},
\label{eqn:dynfric}
\end{equation}
where $M$ is the mass of a satellite moving with velocity $\mathbf{v}$
in the density field $\rho$ of some host, $X = v / (\sqrt{2}\sigma)$
where $\sigma$ is the local 1D velocity dispersion of the host, and
$\ln \Lambda$ is the Coulomb logarithm. This quantity is taken as
\begin{equation}
\ln \Lambda = \ln \left( \frac{r}{\epsilon} \right),
\end{equation}
where we have followed \citet{Ha03} and used the instantaneous
separation $r$ in place of the maximum impact parameter originally
advocated by Chandrasekhar.  \citet{Ha03} showed that this choice gave
a better reproduction of the orbital decay time-scale.

From this set of initial conditions, we run a live disruption forward
-- which of course self-consistently allows for the effects of
dynamical friction -- and see if it approaches the target coordinates
at the Crossover Point closely enough.  If not, we update $\epsilon$ and
re-run. Typically, 2 such iterations are needed. The tolerance with
which the phase space coordinates are matched at the Crossover Point
is 1 kpc and 5 kms$^{-1}$. We explore a grid of Sgr masses in the range
$10^{10}$ to $10^{11}$ and the choose $\rs$ from the
mass-concentration relation, whilst the length $\epsilon$ is in the
range 6 to 12. 

Finally, we need to extract the velocity dispersion profiles from the
simulations. We transform the Galactocentric coordinates of the
particles into heliocentric ones, and thence obtain
$(\tilde{\Lambda}_\odot, \tilde{B}_\odot)$ for particles. In each of
the bins in $\tilde{\Lambda}_\odot$, we fit a Gaussian to extract the
velocity dispersion.  This works well for the trailing arm, but the
leading arm is messy. The leading debris shows several distinct
particle groups in the space of ($v_{\rm GSR}, \tilde{\Lambda}_\odot$)
belonging to individual stripping epochs. To account for that, for the
leading debris analysis, we fit 2 Gaussians, at each
$\tilde{\Lambda}_\odot$.

\begin{figure*}
\centering
\includegraphics[width=0.49\textwidth]{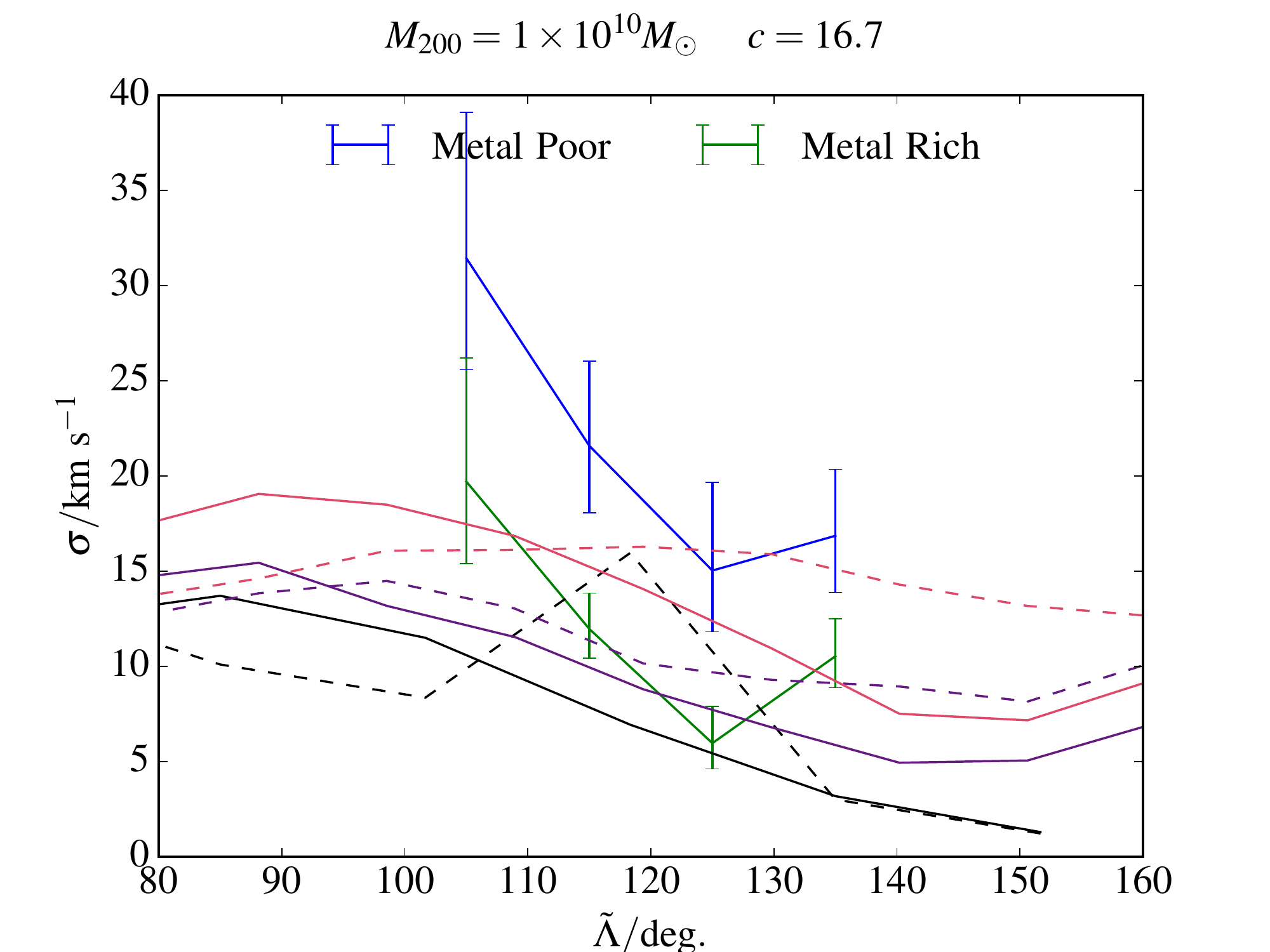}
\includegraphics[width=0.49\textwidth]{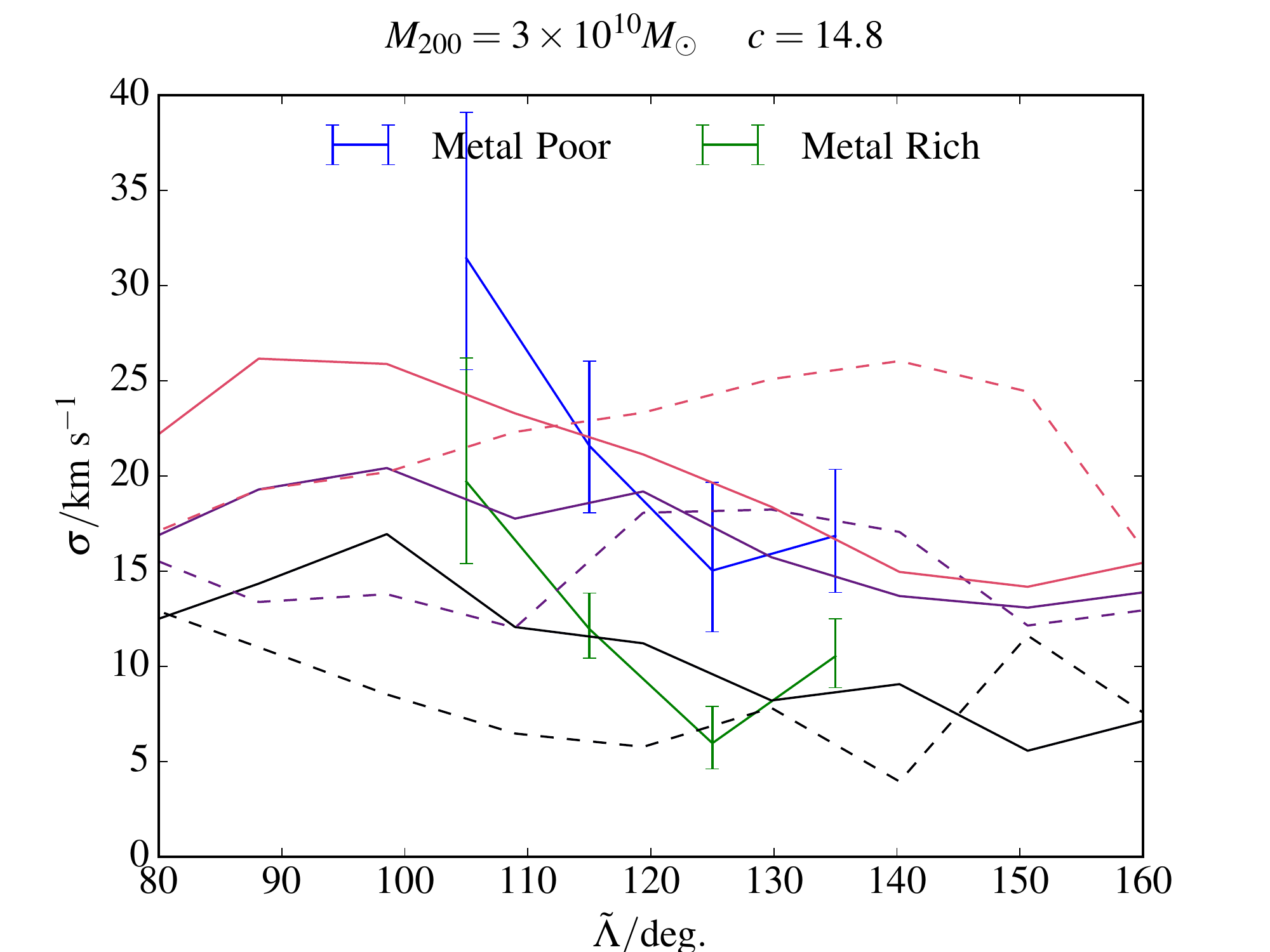}
\includegraphics[width=0.49\textwidth]{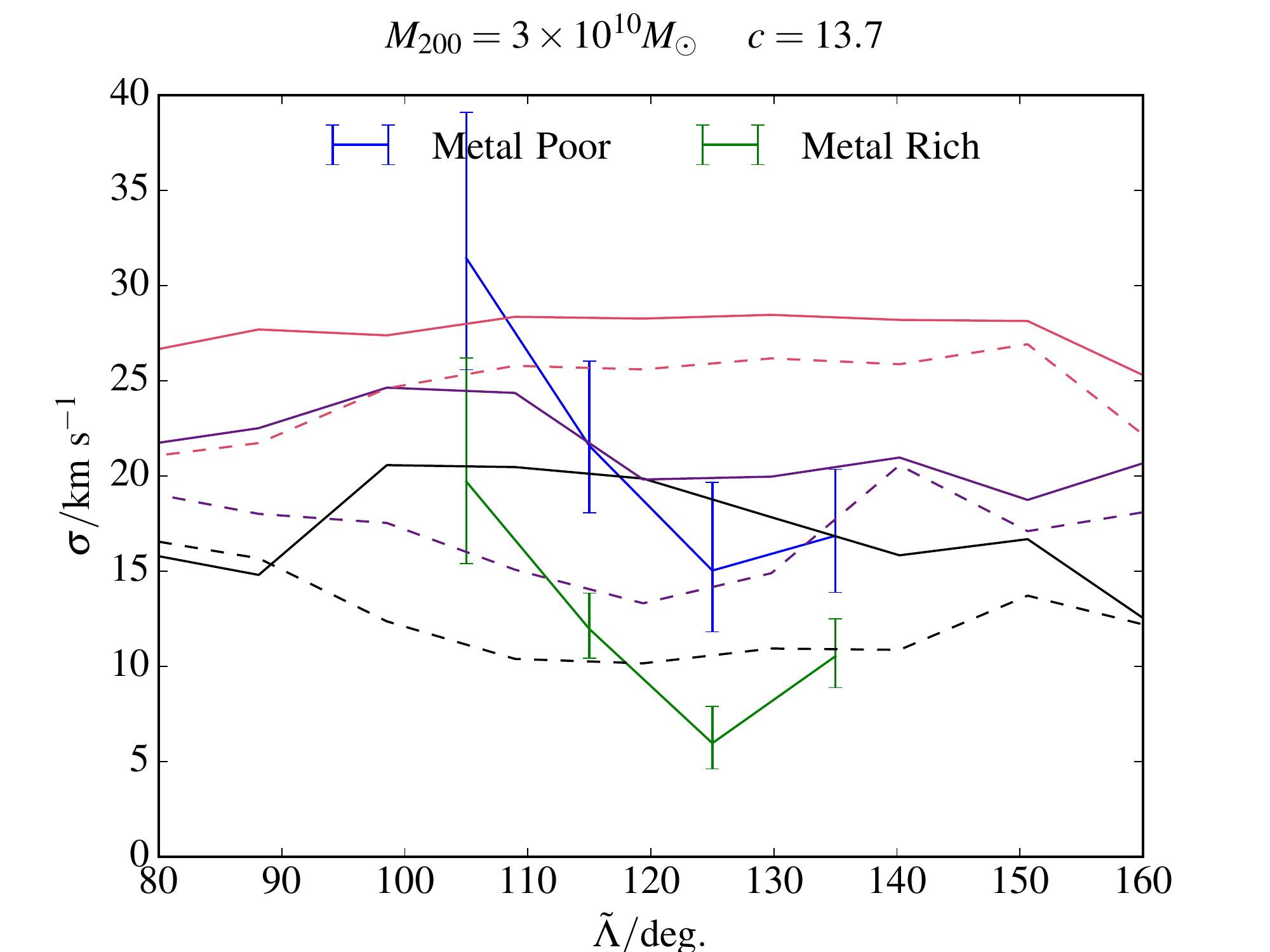}
\includegraphics[width=0.49\textwidth]{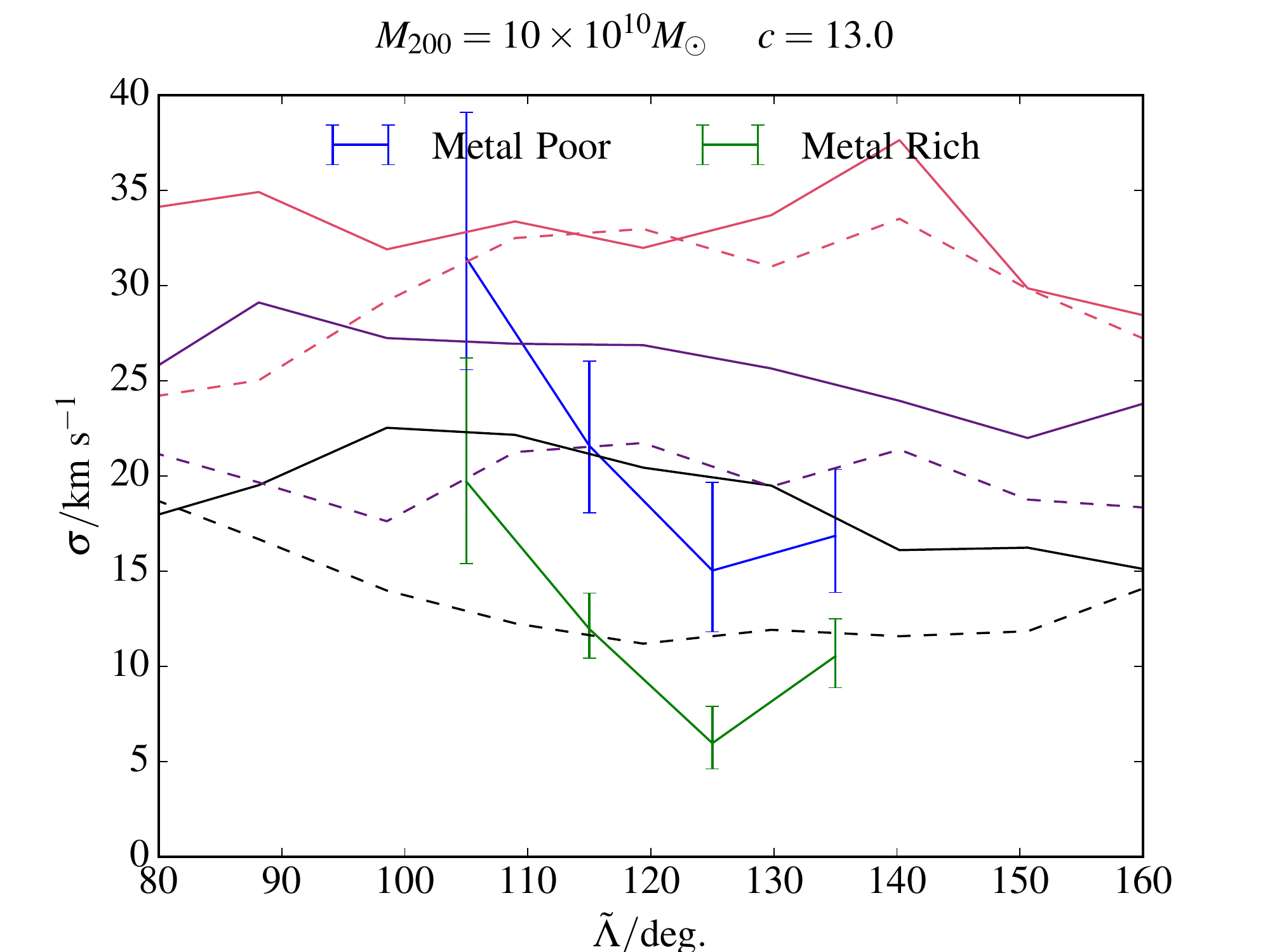}
\caption{As Fig.~\ref{fig:trailvdisps}, but for the leading stream. In
  the simulation, material stripped at different pericentric passages
  is distinguished by different line type. While the phase-space
  behaviour of the leading debris is markedly more complex as compared
  to the trailing one, it is re-assuring to see that our models
  reproduce the elevated velocity dispersions observed in the leading
  tail.}
\label{fig:leadvdisps}
\end{figure*}


\subsection{Results of the N-body experiments}

Fig.~\ref{fig:orbitSgr} displays the evolution of the Sgr orbit and
the mass in its dark matter halo contained within the tidal
radius. This shows the gamut of choices first articulated by
\citet{jiang2000} and then amplified by \citet{Martin2010}. The Sgr
could easily once have been as massive as $10^{11} M_\odot$, in which
case it would have had an apocentric distance of $\sim 110$ kpc as
recently as 5 Gyr ago. However, if its initial dark matter mass is
reduced to $10^{10} M_\odot$, then it must have been correspondingly
closer at $\sim 70$ kpc. Note that the Crossover Point - in essence,
the first apocentre in the \citet{LM2010} potential - is reached at
slightly different epochs by progenitors with different masses. This
is unsurprising, given different amount of dynamical friction, and
hence, orbital evolution. The right panel illustrates the ruthlessness
of the stripping process. The more massive the halo, the greater the
efficacy of the stripping so that all the models end up with modern
day remnants with dark matter mass $\lesssim 10^9 M_\odot$. There is a
very small discontinuity in the run of the progenitor's bound mass at
the Crossover Point. This is due to the fact the \citet{LM2010} MW is
slightly denser, thus giving a smaller tidal radius for the Sgr dwarf
at the same Galact-centric distance compared to the TF model.

Fig.~\ref{fig:sim_properties} shows the behaviour of the luminous mass
and velocity dispresion.  Now, we have the freedom to vary not just
the mass of the progenitor but also the embeddedness of the stellar
light. Interestingly, the present day velocity dispersion is largely
independent of the embeddedness as the curves in the upper panel
converge at late times ($\sim 5$ Gyr). If the progenitor mass is
$10^{10} M_\odot$, then the velocity dispersion of the stars in the
remnant is under 10 kms$^{-1}$. This rises to $\sim 15$ kms$^{-1}$ if
the progenitor mass is $6 \times 10^{10} M_\odot$. In comparison, the
first measurement of the velocity dispersion of giant stars in the Sgr
remnant was by \citet{Ib97}, who found $11.4 \pm 0.7$ kms$^{-1}$.
Later investigations confirmed that the Sgr remnant has a flat
velocity dispersion profile with an amplitude of $\sim 14$ kms$^{-1}$
out to $10^\circ$ (see Fig. 11 of \citet{frinchaboy2012}). However,
caution is needed in comparing the velocity dispersion of the remnant
in the simulations with the observations. In the initial set-up, the
King profile was taken as a tracer density in eq.~(\ref{eq:eddington})
and this may not be justifiable in the centre.  Our approach is fine
for the outer regions which form the tails, and so we now turn to
these.

Please note that caution is required when comparing the results of our
simulations directly to the data in hand. For example, it appears that
both in terms of the DM mass shown in Fig.~\ref{fig:orbitSgr} and the
stellar mass displayed in the bottom row of
Fig.~\ref{fig:sim_properties}, the final state of the remnant is
more depleted than it is implied by observations. More precisely, in
the simulations, the final mass is lower than $10^8$ M$_{\odot}$ and
the remnant's luminosity today is less than the half of the
original stellar mass, while according to \citet{Majewski2013} the
remnant's mass should exceed $10^8$ M$_{\odot}$, and the Sgr has
probably only lost $50\%$ of its stars \citep[see
  e.g.][]{Martin2010}.  However, while our choice of most of the
model parameters is well motivated, plenty of freedom is still
allowed. For example, the disruption time is probably uncertain at the
level of 1 to 2 Gyrs. As such, the state of the progenitor 1 Gyr
before the end of the simulation (as permitted by the current level of
uncertainty) would match the observational constraints better.

Fig.~\ref{fig:trailvdisps} shows the run of velocity dispersions in
the trailing tail, compared to the data extracted in Section 2.  In
the surviving dSphs, such as Scultor or Fornax, the half-light radius
of the metal-rich population is a factor of $\sim 2$ smaller than the
metal-poor~\citep[see e.g.,][]{Am12,Ag12}. The coloured lines refer to
stellar populations in which the ratio $\rK/\rs$ differs by successive
factors of 2. So, it is reasonable to expect a good match between any
successive pair and the data. Such is the case for the heavier mass
progenitors with total dark halo mass $\gtrsim 6 \times 10^{10}
M_\odot$. In both of the lower panels of Fig.~\ref{fig:trailvdisps},
the match is encouraging especially given the limited number of models
available to span the large parameter space.  Progenitor masses
$\lesssim 3 \times 10^{10} M_\odot$ are unable to provide the hot
metal-poor population. The trailing tail provides a reasonably clean
test, but the picture is slightly more muddled when we turn to the
leading tail in Fig.~\ref{fig:leadvdisps}. Here, the two massive
progenitors with mass $6 \times 10^{10} M_\odot$ and $10 \times
10^{10} M_\odot$, match the data in all $\tilde{\Lambda}_\odot$ bins
in agreement with the trailing tail results. Additionally, in some
(but not all) locations along the leading tail, the progenitor with $3
\times 10^{10} M_\odot$ provides a reasonable fit. Note, however, that
the leading tail simulations are significantly more cumbersome to
interpret, as the line-of-sight velocity distribution is clearly
multi-modal. This is evidenced by different behaviour of the solid and
dashed lines representing the run of the velocity dispersion in each
sub-component in the simulated trailing tail.

\section{Discussion and Conclusions}
\label{sec:results}

Multiple populations in dwarf spheroidals are known to be common.
Dwarf extended star formation histories naturally produce younger,
colder, more metal-rich populations and older, hotter, more metal-poor
populations. According to this picture, we might expect the stripping
process to remove predominantly metal-poor stars at first, but with an
increasing admixture of metal-rich population as time goes by. Just as
the multiple populations in dwarf spheroidals provide powerful
constraints on the potential, so the multiple populations in the Sgr
stream may be used to measure the mass of the Sgr progenitor's halo.

This paper has identified multiple sub-populations in the Sagittarius
(Sgr) stream using the SDSS/SEGUE spectroscopic dataset. The
metallicity distribution functions (MDFs) of both the leading and
trailing arms have been decomposed into Gaussians representing
metal-rich and metal-poor populations respectively. In this procedure,
contamination of the sample by Milky Way halo stars is accounted for
by constructing a data-driven empirical MDF of the halo. We find that
the MDF of the Sgr trailing stream is well represented by two
Gaussians with means and dispersions $(\mu^{[Fe/H]}_{\rm
  ST1},\sigma^{[Fe/H]}_{\rm ST1})=(-1.33, 0.27)$ and
$(\mu^{[Fe/H]}_{\rm ST2},\sigma^{[Fe/H]}_{\rm ST2})=(-0.74, 0.18)$
dex.  In agreement with earlier studies~\citep{Sbordone2007,
  Koposov2013, deBoer2014}, the trailing stream has a substantial
metal-rich population. The leading stream's MDF can also be decomposed
into two Gaussians with means and dispersions $(\mu^{[Fe/H]}_{\rm
  SL1},\sigma^{[Fe/H]}_{\rm SL1})=(-1.39, 0.22)$ and
$(\mu^{[Fe/H]}_{\rm SL2},\sigma^{[Fe/H]}_{\rm SL2})=(-1.00, 0.34)$. We
find that - as seen by the SDSS/SEGUE - the metal-rich population in
the leading stream is less prominent.

We link the different metallicity components to their kinematics by
determining the mean velocity and dispersion of the sub-populations as
a function of longitude along the stream. In all the longitude bins
and for both tails, the metal-poor population has a larger velocity
dispersion than the metal-rich. For the trailing tail, the metal-rich
sub-population has kinematics well approximated by a Gaussian with
$\sigma^{\rm v}_{\rm ST2} \sim 8$ km s$^{-1}$, in accord with the
results of \citet[][]{Monaco2007} using high resolution spectroscopy
of M-giant stars. The metal-poor component is hotter with $\sigma^{\rm
  v}_{\rm ST2} \sim 13$ km s$^{-1}$, similar to the value found
by~\citet{Koposov2013}. These values show modest variation with
longitude.  However, the dispersion of the Sgr leading stream depends
more strongly on longitude.  We find $15$ kms$^{-1}< \sigma^{\rm
  v}_{\rm SL1} < 30 $kms$^{-1}$ for the metal-poor sub-population, and
$6$ kms$^{-1}< \sigma^{\rm v}_{\rm SL2} < 20 $kms$^{-1}$ for the
metal-rich. Some of the variation is produced by projection effects,
though some is produced by feathering or streamlets and some by
orbital evolution of the stream in the host potential.

As the velocity dispersions for both populations in both tails can be
measured across a swathe of longitudes, this gives a powerful
constraint on the mass of the progenitor. By carrying out simulations
of the disruption of Sgr in the Milky Way, we find that the starting
Sgr mass has to be $\gtrsim 6 \times 10^{10} M_\odot$. If the mass is
less than this, then the velocity dispersion of the simulated
metal-poor stream is too low, irrespective of the embeddedness of the
population in the original progenitor.

Such a high mass for the Sgr is consistent with a number of other
lines of evidence.  For example, models of \citet{GarrisonKimmel2016}
would predict a mass consistent with $10^{11}$ for Sgr when using the
determination of the total progenitor luminosity of $\sim 10^8
L_\odot$ from \cite{Martin2010}. Moreover, the metallicity of the
``knee'' in the distribution of the abundance of $\alpha$ elements at
given [Fe/H] most likely - albeit in an indirect way - correlates with
the mass of the dwarf galaxy. As shown by \citet{deBoer2014}, the
$\alpha$-knee in the Sgr dSph is just under that of the LMC, thus
confirming the Sgr's ranking as the next most massive dwarf after the
Magellanic Clouds. Note that the LMC itself has been recently
estimated to to have at least $10^{11}$M$_{\odot}$ in DM, or perhaps
even more \citep[see e.g.][]{Jorge2016,Jethwa2016}. The number of
globular clusters (or, more precisely, their total mass) the galaxy
hosts is also known to scale with its DM mass
\citep[][]{Hudson2014}. There are probably at least 8 GCs associated
with the Sgr dwarf according to \citet{LM_gcs}, i.e. the same number
the SMC hosts and half of the known contingent of the LMC.

The gap between the values of the progenitor's mass suggested by our
study and those assumed so far as part of the disruption modelling
\citep[see e.g.][]{LM2010,Gibbons2014} is at least two orders of
magnitude. The difference might actually be a lot less frightening as
the value relevant for the stream centroid analysis is that at the
onset of the stripping of the stellar component. As evidenced in
Fig.~\ref{fig:sim_properties}, the disruption of the luminous portion
of the dwarf starts in earnest at least $\sim$1 Gyr after passing the
Crossover Point. By then, more than 90$\%$ of the DM mass is gone,
thus bringing the total progenitor's mass much closer to
$10^9$M$_{\odot}$. Note that the DM the Sgr has shed might play an
important role in the reconstruction of the MW gravitational
potential: in the end, an amount of the DM dumped onto the Galaxy in
the plane perpendicular to the Galactic disk is at least the mass of
the disk itself! The Sgr DM debris could represent the most
significant halo sub-structure in the Galaxy today, thus affecting the
interpretation the DM direct searches \citep[see e.g.][]{Freese2013,
  Ohare2014}.

Our work also has implications for the Milky Way warp. The idea that
the tidal interaction of large satellite galaxies may excite warps has
been proposed before, often with the suggestion that the Magellanic
Clouds may be the culprit~\citep[e.g.,][]{Hu69,We06}. This is unlikely
if the Magellanics are on first infall, which is the most likely
interpretation of the proper motion data~\citep{Be07,Ka13}. If so,
then another thuggish intruder into the Milky Way halo needs to be
sought. A corollary of making the Sgr much more massive is that the
havoc it wrought in the Milky Way disk is much more extensive. So, the
Sgr must surely be the major suspect into ongoing enquiries as to the
cause of the warp.

\section*{Acknowledgements}
SG thanks the Science and Technology Facilities Council (STFC) for the
award of a studentship.  The research leading to these results has
received funding from the European Research Council under the European
Union's Seventh Framework Programme (FP/2007-2013) / ERC Grant
Agreement no. 308024.
\label{lastpage}

\bibliographystyle{mn2e}
\bibliography{refs}

\end{document}

%% file: journal_defs.tex
\long\def\symbolfootnote[#1]#2{\begingroup%
\def\thefootnote{\fnsymbol{footnote}}\footnote[#1]{#2}\endgroup} 

%% file: my_macros.tex
\usepackage{ifthen}

\newcommand{\dd}{ {\rm d} }                                

\newcommand{\ddv}[3][1]{                                   
            \ifthenelse{\equal{#1}{1}}                     
                       {\frac{\dd #2}{\dd #3}}             
                       {\frac{\dd^{#1} #2}{\dd #3^{#1}}}}  

\newcommand{\pddv}[3][1]{                                  
            \ifthenelse{\equal{#1}{1}}
                       {\frac{\partial #2}{\partial #3}}
                       {\frac{\partial^{#1} #2}{\partial #3^{#1}}}}

\newcommand{\kpc}{\ensuremath{\rm kpc}}

\newcommand{\kms}{\ensuremath{\rm km \, s^{-1}}}

\newcommand{\K}{\ensuremath{\rm K}}
